\documentclass[12pt]{elsarticle}

\usepackage{amssymb}

\journal{Computer Physics Communications}

\usepackage{amsmath}

\usepackage{mathtools}
\usepackage{amsthm}
\usepackage{subfigure}
\usepackage{xspace}
\usepackage{fullpage}
\usepackage{comment}

\usepackage{hyperref}

\usepackage{tikz}

\newlength{\colwidth}
\usetikzlibrary{calc}
\usetikzlibrary{calc,trees,positioning,arrows,chains,shapes.geometric,%
    decorations.pathreplacing,decorations.pathmorphing,shapes,%
    matrix,shapes.symbols}
    
\tikzset{nodeStyle/.style = {circle,draw,minimum size=30pt}}
\tikzset{arrowStyle/.style = {-latex}}
\tikzset{nodeStyle/.style = {circle,draw,minimum size=2.5em}}
\tikzset{EdgeStyle/.style = { thick,-triangle 45}}	
\tikzset{arrowEdge/.style = { thick}}	
\tikzset{additionalEdgeStyle/.style = {decorate,decoration=snake,draw=red}}
\tikzset{boxStyle/.style = {thick, dashed}}
\usepackage{listings}
\usepackage{courier}

\lstdefinelanguage{julia}
  {morekeywords={
    typeof,typemin,typemax,eps,zero,one,Inf,NaN,ones,eye, % numerics
    Int,Int64,Float64,Int32,Float32,Complex128,Complex64,BigInt,BigFloat,Dict,Graph,Array,%
    begin,end,function,using,if,elseif,else,while,for,% control flow
    try,catch,error,throw,yieldto, % control flow
    Pkg, add, update, clone, names, %package management
    println, full, push, !, collect, sum, diag, ceil, plot, count, length, 
    eigvales, abs, Base, vec % varia
  },
  sensitive=true,
  morecomment=[l]{\#},
  morecomment=[s]{"""}{"""},
  morestring=[b]',
  keywordstyle=, % use the standard font for keywords 
}

\newcommand{\julia}[1]{\lstinline{#1}}
\newcommand{\pkgName}[1]{\textsc{#1}}
\newcommand{\package}{\pkgName{QSWalk.jl}\xspace}
\newcommand{\Julia}{\textit{Julia}\xspace}
\newcommand{\Jupyter}{\pkgName{Jupyter}\xspace}
\newcommand{\MMA}{\textit{Mathematica}\xspace}
\newcommand{\WMMA}{\textit{Wolfram Mathematica}\xspace}

\newcommand{\file}[1]{\lstinline[language=csh]{#1}}
\newcommand{\mmaCode}[1]{\texttt{#1}}

\newcommand{\ket}[1]{\ensuremath{|#1\rangle}}
\newcommand{\bra}[1]{\ensuremath{\langle#1|}}
\newcommand{\ketbra}[2]{\ensuremath{\ket{#1}\bra{#2}}}

\newcommand{\LL}{\mathcal{L}}

\newcommand{\Id}{\mathrm{I}}
\newcommand{\ii}{\mathrm{i}}
\newcommand{\dd}{\mathrm{d}}

\newcommand{\kron}{\otimes}
\newcommand{\con}[1]{\bar{#1}}
\newcommand{\vecc}[1]{ | #1 \rangle \rangle}

\newcommand{\outdeg}{\operatorname{outdeg}}

\newcommand{\spann}{\operatorname{span}}
\newcommand{\ie}{ie.\xspace}
\newcommand{\eg}{eg.\xspace}

\newcommand{\docName}{paper\xspace}

\newcounter{note}
\begin{document}
\begin{frontmatter}



\title{\package{}: Julia package for  quantum stochastic walks analysis}


\author[a,b]{Adam Glos\corref{author}}
\author[a]{Jaros\l{}aw Adam Miszczak}
\author[a,b]{Mateusz Ostaszewski}

\cortext[author] {Corresponding author.\\\textit{E-mail address:} aglos@iitis.pl}
\address[a]{Institute of Theoretical and Applied Informatics, Polish Academy of 	Sciences \\Ba{\l}tycka 5, 44-100 Gliwice, Poland}
\address[b]{Institute of Informatics, Silesian University of Technology \\ Akademicka 16, 44-100 Gliwice, Poland}
	
\begin{abstract}
The \docName{} describes \package{} package for Julia programming
language, developed for the purpose of simulating the evolution of open quantum systems. The
package enables the study of quantum procedures developed using stochastic
quantum walks on arbitrary directed graphs. We provide a detailed description of
the implemented functions, along with a number of usage examples. The package is
compared with the existing software offering a similar functionality.\\[7pt]
		
%		A submitted program is expected to be of benefit to other physicists or physical chemists, or be an exemplar of good programming practice, or illustrate new or novel programming techniques which are of importance to some branch of computational physics or physical chemistry.
%		
%		Acceptable program descriptions can take different forms. The following Long Write-Up structure is a suggested structure but it is not obligatory. Actual structure will depend on the length of the program, the extent to which the algorithms or software have already been described in literature, and the detail provided in the user manual.
%		
%		Your manuscript and figure sources should be submitted through the Elsevier Editorial System (EES) by using the online submission tool at \\
%		http://www.ees.elsevier.com/cpc.
		
%		In addition to the manuscript you must supply: the program source code; job control scripts, where applicable; a README file giving the names and a brief description of all the files that make up the package and clear instructions on the installation and execution of the program; sample input and output data for at least one comprehensive test run; and, where appropriate, a user manual. These should be sent, via email as a compressed archive file, to the CPC Program Librarian at cpc@qub.ac.uk.
		
\end{abstract}
	
	
\end{frontmatter}

{\bf PROGRAM SUMMARY}
%Delete as appropriate.

\begin{small}
	\noindent
	{\em Program Title:} \package                                         \\
	{\em Licensing provisions:} MIT                                   \\
	{\em Distribution format:} Source code available at \url{https://github.com/QuantumWalks/QSWalk.jl}\\
	{\em Programming language:} Julia                                   \\
%	{\em Supplementary material:}                                 \\
	% Fill in if necessary, otherwise leave out.
%	{\em Journal reference of previous version:}                  \\
	%Only required for a New Version summary, otherwise leave out.
%	{\em Does the new version supersede the previous version?:}   \\
	%Only required for a New Version summary, otherwise leave out.
%	{\em Reasons for the new version:}\\
	%Only required for a New Version summary, otherwise leave out.
%	{\em Summary of revisions:}*\\
	%Only required for a New Version summary, otherwise leave out.
%	
	{\em Nature of problem:} The package implements functions for simulating
	quantum stochastic walks, including local regime, global regime, and
	nonmoralizing global regime \cite{domino2017superdiffusive_summary}. It can be used
	for arbitrary quantum continuous evolution based on GKSL master equation on
	arbitrary graphs. \\
	%Describe the nature of the problem here. \\
	{\em Solution method:} We utilize Expokit routines for fast sparse matrix
	exponentials on vectors. For dense matrices exponentiation is computed
	separately, which is faster for small matrices. \\
	%Describe the method solution here.
	{\em Restrictions:}
	Currently package requires Julia v0.6.\\

%	\\
%	* Items marked with an asterisk are only required for new versions
%	of programs previously published in the CPC Program Library.\\
\end{small}

%%%%%%%%%%%%%%%%%%%%%%%%%%%%%%%%%%%%%%%%%%%%%%%%%%%%%%%%%%%%%%%%%%%%%%%%%%%%%%%%
\section{Introduction}
%%%%%%%%%%%%%%%%%%%%%%%%%%%%%%%%%%%%%%%%%%%%%%%%%%%%%%%%%%%%%%%%%%%%%%%%%%%%%%%%
During the last few years \Julia\ programming language
\cite{julia-docs,bezanson2017julia} became highly popular in the scientific
computing community because of its dynamical style of programming, combined with
strong typing and high efficiency. As such it provides a very attractive
platform for simulating models used in quantum information processing.

In this \docName we focus on the particular model used to study the dynamics of
quantum systems, namely quantum stochastic walks. The model is used to study
quantum-to-classical transition and to probe the influence of quantum evolution
on the efficiency of information processing \cite{domino2016properties}. Quantum
stochastic walks have been widely analysed in context of propagation
\cite{domino2016properties,glos2017limit,liu_continuous-time_2016,domino2017superdiffusive,bringuier_central_2016}, application in computer science \cite{sanchez-burillo_quantum_2012,falloon2017qswalk,loke2017comparing} and physics \cite{caruso2009highly,mohseni2008environment,rebentrost2009environment}. In particular, new model of fast-propagating quantum walk on arbitrary directed graph was developed \cite{glos2017limit,domino2017superdiffusive}. To this end, we describe a package for \Julia\ programming language developed for the purpose of simulating and analyzing quantum stochastic walks~\cite{whitfield2010quantum} and we compare its efficiency with the existing software.

 %o fizyce (zastooswania)

The main goal of this paper is to describe a package \package for \Julia
programming language which enables high-performance analysis of quantum
stochastic walks. There are three main advantages of the package over
the existing software. First, it enables the simulation of quantum stochastic
walks analysis in both local and global regimes \cite{glos2017limit}. Second, it
is the first package providing the implementation of the nonmoralizing
evolution. Thus, it enables the simulation of fast continuous-time quantum walk
model on arbitrary directed graph \cite{domino2017superdiffusive}. Finally, the
package can be used to perform numerical experiments for very large
graphs thanks to sparse matrices implementation used. This is especially
important for investigating the statistical properties of complex graphs.

\paragraph{Comparison with the existing software} Many software packages for
simulating quantum computing~\cite{quantiki-list}, including some focused on
various models of discrete quantum walks
\cite{marquezino2008qwalk,berry2011qwviz} and continuous-time quantum
walks~\cite{falloon2017qswalk,izaac2015pyctqw} are available. Most of them are
focused on the discrete-time evolution, restricted to the unitary operations and
undirected graphs. The first exception is \pkgName{QSWalk.m} package
\cite{falloon2017qswalk} for \WMMA, which provides implementation of a quantum
stochastic walk model. Package \pkgName{QSWalk.m} utilizes \mmaCode{MatrixExp}
function and sparse representation of superoperators for simulating the
evolution of quantum stochastic walks. It provides the basic functionality
required to simulate quantum stochastic walks. However, it has limitations
related to the performance of simulation of high dimensional systems, see
Sec.~\ref{sec:performance}. In contrast, the package described in this \docName
enables the simulation of much larger systems. This is crucial for the analysis
of complex networks and the analysis of the convergence on large graphs. It can
be also utilized to calculate the average transfer time, which is a relevant
measure for a quantum transport process \cite{mohseni2008environment,rebentrost2009environment}.

 In the context of physics application, in many cases it is necessary to analyze
the convergence properties of the evolution. For some cases convergence
criteria suitable for quantum reservoir engineering were given
\cite{schirmer2010stabilizing}. However, such criteria cannot be always
obtained, hence spectral analysis of the evolution superoperator is necessary.
The \package package facilitates the analysis of the quantum stochastic walks
by providing access to the functions for constructing generators of the
dynamical subgroup and the superoperators. Spectral analysis of evolution
superoperator, enabled by our package, can be used for verification of the
convergence criteria. In case of noise-assisted transport
\cite{caruso2009highly}, this can be used to derive stationary state, and by
this a total amount of the excitation that is transferred to the sink.
Furthermore, superoperator eigenvalues with small magnitude determine the
convergence rate. 
 
Moreover, the package provides the implementation of non-moralizing quantum walk
model~\cite{domino2017superdiffusive}. This type of walk enables fast
digraph-preserving evolution. None of the existing software package provides the
implementation of this model.

Additionally, \Julia is an open-source programming language. It can be used for
free for academic as well as commercial purposes. It enables a seamless
utilization of parallel computing for numerical analysis, which facilitates the
analysis of large complex networks.

%nasza paczka spełnia oczekiwania

%%%%%%%%%%%%%%%%%%%%%%%%%%%%%%%%%%%%%%%%%%%%%%%%%%%%%%%%%%%%%%%%%%%%%%%%%%%%%%%%
\section{Theoretical background}\label{sec:theory}
%%%%%%%%%%%%%%%%%%%%%%%%%%%%%%%%%%%%%%%%%%%%%%%%%%%%%%%%%%%%%%%%%%%%%%%%%%%%%%%%
Before describing the functions provided by \package package, we review
essential theoretical concepts used to define \emph{quantum stochastic walks}.
This model of quantum evolution defines the family of quantum processes in the
discrete state space and with the continuous time parameter.

First, we recall the definition of the GKSL master equation, which describes the
continuous evolution of the open quantum system. Next, we introduce quantum
stochastic walks and briefly describe the local and global interaction regimes.
Finally, we describe spontaneous moralization which is a side-effect of using
global interaction model on directed graphs.

%%%%%%%%%%%%%%%%%%%%%%%%%%%%%%%%%%%%%%%%%%%%%%%%%%%%%%%%%%%%%%%%%%%%%%%%%%%%%%%%
\subsection{GKSL master equation}
%%%%%%%%%%%%%%%%%%%%%%%%%%%%%%%%%%%%%%%%%%%%%%%%%%%%%%%%%%%%%%%%%%%%%%%%%%%%%%%%

%Let us denote by $\States{N}$ the set of density matrices of dimension $N$, \ie\ the subset of $N\times N$ matrices such that its elements are positive semi-definite and have unit trace.
The starting point for introducing quantum stochastic walks is the
Gorini-Kossakowski-Sudarshan-Lindblad (GKSL) master equation
\cite{ gorini1976completely, kossakowski1972quantum, lindblad1976generators}
\begin{equation}
	\frac{\dd}{\dd t}\varrho = -\ii [H, \varrho] + 
	\frac{1}{2} \sum_{L\in 
	\LL} \left(2  L \varrho 
	L^\dagger -   L^\dagger L \varrho -  \varrho L^\dagger L  \right), 
	\label{eq:GKSL-master-equation}
\end{equation}
where $H$ is the Hamiltonian, which describes the evolution of the closed
system, and $\LL$ is the collection of Lindblad operators, which describes the
evolution of the open system. Operator $\varrho$ is the density matrix
representing the state of the system. While the Hamiltonian operator needs to be
Hermitian, there is no general requirement on Lindblad operators $L\in\LL$. This
master equation describes the general continuous evolution of mixed quantum
states.

In the case where $H$ and $\LL$ do not depend on time, we say that
Eq.~\eqref{eq:GKSL-master-equation} describes the Markovian evolution of the
system. Henceforth, we can solve the differential equation analytically. If we
choose initial state $\varrho(0)$, then
\begin{equation}
\vecc{\varrho(t)} = \exp(t F) \vecc{\varrho(0)}, \label{eq:evolution}
\end{equation}
where
\begin{equation}
F = \left(H \kron \Id - \Id \kron \con H \right) + 
\sum_{L\in\LL} \left ( L \kron \con L - \frac{1}{2}\left( L^\dagger L \kron 
\Id 
+ \Id \kron L^\dagger  L\right )\right 
)\label{eq:evolve-generator}
\end{equation}
and $\vecc{\,\cdot\,}$ denotes the vectorization of the
matrix~\cite{miszczak2011singular}. We call $F$ the \emph{evolution generator} and
corresponding $\exp(t F)$ a \emph{superoperator}. 
%Those equations are basic for our algorithm construction.

%Another equivalent form of the Eq.~\eqref{eq:GKSL-master-equation} is in
%particularly interesting.
Since $H$ and $\LL$ describe closed and open system evolution respectively, it
is interesting how the change of impact of these parts affects the evolution.
Hence we add smoothing parameter $\omega\in[0,1]$ and change
Eq.~\eqref{eq:GKSL-master-equation} into
\begin{equation}
\frac{\dd }{\dd t}\varrho = -\ii (1-\omega)[H, \varrho] + 
\frac{1}{2}\omega \sum_{L\in 
	\LL} \left(2  L \varrho 
L^\dagger -   L^\dagger L \varrho -  \varrho L^\dagger L  \right). \label{eq:evolution-smoothing}
\end{equation}
The evolution generator $F$ and the superoperator $\exp(t F)$ changes
analogically. Note that for $\omega=0$ we recovered the closed system evolution
and for $\omega=1$ we have the purely open evolution.
%Such approach has already been used
%\cite{domino2016properties,domino2017superdiffusive,glos2017limit}.

%%%%%%%%%%%%%%%%%%%%%%%%%%%%%%%%%%%%%%%%%%%%%%%%%%%%%%%%%%%%%%%%%%%%%%%%%%%%%%%%
\subsection{Quantum stochastic walks} 
%%%%%%%%%%%%%%%%%%%%%%%%%%%%%%%%%%%%%%%%%%%%%%%%%%%%%%%%%%%%%%%%%%%%%%%%%%%%%%%%
The GKSL master equation describes the general form of the evolution of the open
quantum system. If one introduces the Hamiltonian and the Lindblad operators
using graphs as the underlying structure, the resulting process can be
understood as the walk evolution.

In this manner, GKSL master equation was used for defining quantum stochastic
walks. This model of evolution provides the generalization of both classical
random walks and quantum walks \cite{whitfield2010quantum}. In this case $H$ and
$\LL$ in Eq.~\eqref{eq:GKSL-master-equation} are constructed using the graph
structure. However, one may verify that the choice of Lindblad operators may be
non-unique \cite{domino2016properties,whitfield2010quantum}.

Let us suppose a directed graph $G=(V,A)$ is given, with $V=\{v_1,\dots,v_n\}$.
By $\bar G = (V,E)$ we denote the underlying undirected graph of $G$. Two main
variants of quantum stochastic walks can be distinguished. In the \emph{local
environment interaction case} each Lindblad operator corresponds to a single
edge, $ \LL^l = \{\ketbra{w}{v}\colon (v,w)\in A\}$. On the other hand, in the
\emph{global environment interaction case} we choose a single Lindblad operator
$\LL^g=\{\sum_{(v,w)\in A }\ketbra{w}{v}\}$. In both models we choose the
Hamiltonian $H$ to be the adjacency matrix of the underlying graph $\bar G$.
Both models were analyzed in context of propagation
\cite{domino2016properties,bringuier_central_2016} and convergence
\cite{glos2017limit,liu_continuous-time_2016,domino2017superdiffusive}.

\begin{figure}\centering
	\begin{tabular}{ccc}
		\begin{tikzpicture}[scale=0.8,node distance=2.5cm,thick]
		
		\node[nodeStyle] (A)  {$v_1$};
		\node[nodeStyle] (C) [below right of=A] {$v_3$};
		\node[nodeStyle] (B) [above right of=C] {$v_2$};
		
		\draw[EdgeStyle] (A) -- (C) node [midway,above] {1} ;
		\draw[EdgeStyle] (B) -- (C) node [midway,above] {1} ;
		
		\end{tikzpicture} & 
		\raisebox{0cm}{\begin{tikzpicture}[scale=0.8,thick]
			
			\coordinate (k1) at (0,0);
			\coordinate (k2) at (1,0);
			\coordinate (k3) at (1,.5);
			\coordinate (k4) at (2,-0.55);
			\coordinate (k5) at (1,-1.5);
			\coordinate (k6) at (1,-1);
			\coordinate (k7) at (0,-1);

			\node[draw=none] (text) at (.8,-.5) {\small GKSL};
			
			\draw[arrowEdge,decorate] (k1) -- (k2) -- (k3) -- (k4) 
			-- (k5) -- 
			(k6) -- 
			(k7) -- 
			(k1);
			
			\end{tikzpicture}} &
		\begin{tikzpicture}[scale=0.8,node distance=2.5cm,thick]
		\tikzstyle{myarrows}=[line width=1mm,-triangle 
		45,postaction={draw, line 
			width=3mm, shorten >=4mm, -}]
		
		\node[nodeStyle] (A)  {$v_1$};
		\node[nodeStyle] (C) [below right of=A] {$v_3$};
		\node[nodeStyle] (B) [above right of=C] {$v_2$};
		
		\draw[EdgeStyle] (A) -- (C) node [midway,above] {1};
		\draw[EdgeStyle] (B) -- (C) node [midway,above] {1};
		
		\draw[additionalEdgeStyle] (A) -- (B) node [midway,above] 
		{\footnotesize 
			$\bra 
			{v_1} L^\dagger L \ket {v_2}\!=\!2$};
		\end{tikzpicture}\\[1ex]
		\begin{tikzpicture}[scale=0.8,thick]
		
		\coordinate (k1) at (-1.2,0);
		\coordinate (k2) at (2.2,0);
		\coordinate (k3) at (2.2,-1.3);
		\coordinate (k4) at (2.8,-1.3);
		\coordinate (k5) at (.5,-2);
		\coordinate (k6) at (-1.8,-1.3);
		\coordinate (k7) at (-1.2,-1.3);

		\node[draw=none] (text) at (0.5,-0.7) {\small CORRECTION};
		
		\draw[arrowEdge,decorate] (k1) -- (k2) -- (k3) -- (k4) -- 
		(k5) -- (k6) 
		-- 
		(k7) -- 
		(k1);
		
		\end{tikzpicture}
		\\[1ex]
		\begin{tikzpicture}[scale=0.8,thick]
		
		\node[nodeStyle] (A) at (0,0) {$v_1$};
		\node[nodeStyle] (C1) at (1,-3) {};
		\node[nodeStyle] (C2) at (3,-3) {};
		\node[nodeStyle] (B) at (4,0) {$v_2$};
		
		\node[] (C3) at (2,-3.5) {$v_3$};
		
		\draw[EdgeStyle] (A) -- (C1) node [midway,right] {1};
		\draw[EdgeStyle] (B) -- (C1) node [midway,right] {1};
		\draw[EdgeStyle] (A) -- (C2) node [midway,right] {1};
		\draw[EdgeStyle] (B) -- (C2) node [midway,right] {-1};
		
		\draw[boxStyle] (0,-2) -- (4,-2) -- (4,-4) -- (0,-4) -- (0,-2);
		
		\end{tikzpicture} & 
		\raisebox{0cm}{\begin{tikzpicture}[scale=0.8]
			
			\coordinate (k1) at (0,0);
			\coordinate (k2) at (1,0);
			\coordinate (k3) at (1,.5);
			\coordinate (k4) at (2,-0.55);
			\coordinate (k5) at (1,-1.5);
			\coordinate (k6) at (1,-1);
			\coordinate (k7) at (0,-1);
			
%			\coordinate (k8) at (0,-3);
			
			\node[draw=none] (text) at (.8,-.5) {\small GKSL};
			
			\draw[arrowEdge,decorate] (k1) -- (k2) -- (k3) -- (k4) 
			-- (k5) -- 
			(k6) -- 
			(k7) -- 
			(k1);
			
			\end{tikzpicture}} &
		\begin{tikzpicture}[scale=0.8]
		
		\node[nodeStyle] (A) at (0,0) {$v_1$};
		\node[nodeStyle] (C1) at (1,-3) {};
		\node[nodeStyle] (C2) at (3,-3) {};
		\node[nodeStyle] (B) at (4,0) {$v_2$};
		
		\node[] (C3) at (2,-3.5) {$v_3$};
		
		\draw[EdgeStyle] (A) -- (C1) node [midway,right] {1};
		\draw[EdgeStyle] (B) -- (C1) node [midway,right] {1};
		\draw[EdgeStyle] (A) -- (C2) node [midway,right] {1};
		\draw[EdgeStyle] (B) -- (C2) node [midway,right] {-1};
		
		\draw[white] (A) -- (B) node [midway,above,black] {\footnotesize					$\bra {v_1} L^\dagger L \ket {v_2}\!=\!0$};
		\draw[boxStyle] (0,-2) -- (4,-2) -- (4,-4) -- (0,-4) -- (0,-2);

		\end{tikzpicture}
%		\begin{tikzpicture}[thick]
%		
%		\node[nodeStyle] (A) at (0,0) {$v_1$};
%		\node[nodeStyle] (C1) at (2,-6) {};
%		\node[nodeStyle] (C2) at (6,-6) {};
%		\node[nodeStyle] (B) at (8,0) {$v_2$};
%		
%		\node[] (C3) at (4,-7.2) {$v_3$};
%		
%		\draw[EdgeStyle] (A) -- (C1) node [midway,right] {1};
%		\draw[EdgeStyle] (B) -- (C1) node [midway,right] {1};
%		\draw[EdgeStyle] (A) -- (C2) node [midway,right] {1};
%		\draw[EdgeStyle] (B) -- (C2) node [midway,right] {-1};
%		
%		
%		
%		
%		\draw[boxStyle] (0,-4) -- (8,-4) -- (8,-8) -- (0,-8) -- 
%		(0,-4);
%		
%		\end{tikzpicture}
	\end{tabular}
    \caption{In the global interaction case the direct application of adjacency
    matrices results in an additional coherent transition. In the example above,
    another connection between $v_1$ and $v_2$ appears. The application of the
    correction procedure described in \cite{domino2017superdiffusive} enables
    the removal of such connections. Note that the correction creates additional
    nodes, which in turn result in the system enlargement.} \label{fig:moralization-example}
\end{figure}
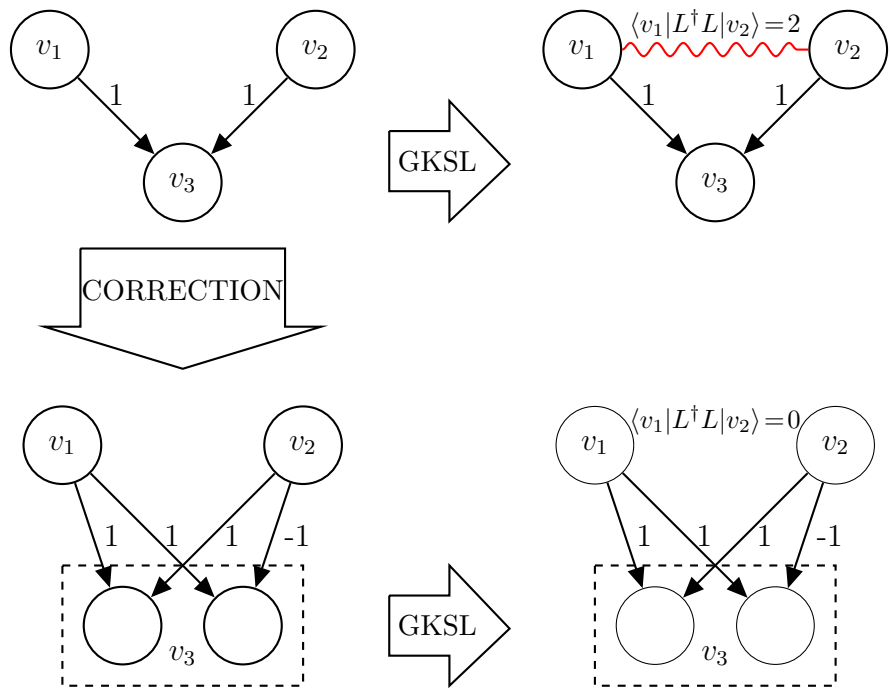

Furthermore, the global interaction was analyzed in the context of graph
preservation. Let us fix global interaction evolution with Hamiltonian $H$ and a
single Lindblad operator $L$. It can be demonstrated
\cite{domino2017superdiffusive} that if two nodes have a common child, then
other amplitude transitions appear between them. This effect is presented in
Fig.~\ref{fig:moralization-example}. To remove such undesirable effect, the
following correction scheme has been proposed.
\begin{enumerate}
	\item For each vertex $v_i$ create $\outdeg(v_i)$ copies, denoted as
	$\tilde v_i^0,\dots,\tilde {v}_i^{\outdeg(v_i)-1}$. If $\outdeg(v_i)=0$, we
	create single copy $\tilde v_i^0$.
	
	\item Construct Hamiltonian $\tilde H$ as 
	\begin{equation}
		\bra{\tilde v_i^k} \tilde H\ket{\tilde v_j^l} = 
        \begin{cases}
        \bra{v_i}H\ket{v_j}, & \{v_i,v_j\}\in E,\\
		0, & \textrm{otherwise}.
		\end{cases}
	\end{equation} 	
	Note that for $\{v_i,v_j\}\in E$ other values may be chosen, but other
	elements should be essentially equal to zero in order to preserve the
	underlying graph structure.
	
	\item Construct Lindblad operator $\tilde L$ as 
	\begin{equation}
	\bra{\tilde v_i^k} \tilde L	\ket{\tilde v_j^l} =
	\bra {v_i}L\ket{v_j} \bra k A_i \ket j.
	\end{equation}
	Here $A_i$ is a matrix which columns are pairwise orthogonal. Such
	definition captures the digraph structure.
	
	\item Construct an additional, acting locally, Hamiltonian $\tilde
	H_{\textrm{loc}}$ for which $\bra{\tilde v_i^k }\tilde H_{\textrm{loc}}
	\ket{\tilde v_j^l}=0$ for $i\neq j$.
\end{enumerate} 

The non-moralizing evolution obtained as the result of the above construction
takes the form
\begin{equation}
\frac{\dd}{\dd t}\tilde \varrho = -\ii (1-\omega)[\tilde H,\tilde  \varrho] + 
\frac{1}{2}\omega \left(-\ii 2[\tilde H_{\textrm{loc}},\tilde \varrho] +2  \tilde L \tilde \varrho 
\tilde L^\dagger -  \tilde  L^\dagger \tilde L \tilde \varrho -  \tilde \varrho \tilde  L^\dagger\tilde  L  \right).\label{eq:evolution-smoothing-nonmoralizing}
\end{equation}

Note that the system 
%used for the he non-moralizing evolution 
enlarges, hence we need to adjust the representation of the
initial state. For this reason, the canonical measurement corresponding to the
vertices of original $G$ graph takes the form
\begin{equation}
p(v_i) =\sum_{k}^{} \bra {\tilde v_i^k} \tilde \varrho_t \ket {\tilde v_i^k},
\end{equation}
where $k$ goes over all copies of $v_i$. This is equivalent to the probability
of measuring any of the copy of the vertex.

Note that for $\LL=\emptyset$ case one can recover the original continuous
quantum walk on mixed states. Hence, the model of quantum stochastic walk
provides a generalization of the quantum walks in closed systems. In this case
another numerical method based on formula
\begin{equation}
\varrho(t)= \exp(-\ii H t)\varrho(0) \exp(\ii H t)
\end{equation}
can be applied. Here the exponentiation is computed for $n\times n$ matrix
instead of $n^2\times n^2$ matrix as in Eq.~\eqref{eq:evolution}.

%%%%%%%%%%%%%%%%%%%%%%%%%%%%%%%%%%%%%%%%%%%%%%%%%%%%%%%%%%%%%%%%%%%%%%%%%%%%%%%%
\section{Package description}\label{sec:package}
%%%%%%%%%%%%%%%%%%%%%%%%%%%%%%%%%%%%%%%%%%%%%%%%%%%%%%%%%%%%%%%%%%%%%%%%%%%%%%%%

In this section we provided the description of \package package and the
dependences required to run simulations created using it. We provide a detailed
description of all its functions present a typical work-flow required to utilize
the package.

\subsection{Installation and recommended packages}
%%%%%%%%%%%%%%%%%%%%%%%%%%%%%%%%%%%%%%%%%%%%%%%%%%%%%%%%%%%%%%%%%%%%%%%%%%%%%%%%
\package{} package can be installed by issuing 
\begin{lstlisting}
julia> Pkg.clone("git://github.com/ZKSI/QSWalk.jl.git")
\end{lstlisting}
within the \Julia interpreter. 
%The package can be updated to the most recent version using 
%\begin{lstlisting}
%julia> Pkg.update("QSWalk")
%\end{lstlisting}
%%%%%%%%%%%%%%%%%%%%%%%%%%%%%%%%%%%%%%%%%%%%%%%%%%%%%%%%%%%%%%%%%%%%%%%%%%%%%%%%
%\subsubsection{Recommended packages}
%%%%%%%%%%%%%%%%%%%%%%%%%%%%%%%%%%%%%%%%%%%%%%%%%%%%%%%%%%%%%%%%%%%%%%%%%%%%%%%%
Package \package{} requires \pkgName{Expokit.jl}  package~\cite{expokit},
which implements the method for calculating matrix exponent described
in~\cite{sidje1998expokit}. Package \pkgName{Expokit.jl} will be installed
automatically during \package installation.

It is recommended to install \pkgName{IJulia.jl} package, which facilitates
interactive development. This package can be installed using
\begin{lstlisting}
julia> Pkg.add("IJulia")
\end{lstlisting}
command. Package \pkgName{IJulia.jl} provides browser-based 
development environment. In order to start \pkgName{IJulia.jl} session one has to 
execute commands
\begin{lstlisting}
julia> using IJulia
julia> notebook(detached=true)
\end{lstlisting}
issued in the interactive \Julia console environment. Directory \file{examples}
in \package distribution contains files with \file{.ipynb} extensions, which can
be loaded using a web browser with running \pkgName{IJulia.jl} session. The
directory contains standard \Julia files with \file{.jl} extension as well.

Additionally, we recommend \pkgName{LightGraphs.jl} package~\cite{lightgraphs},
which provides rich graph related functionality. In particular, it provides the
methods for generating random graphs and their adjacency matrices. The examples
distributed with the package utilize \pkgName{PyPlot.jl} module \cite{pyplot} for
creating plots. The packages are necessary for running the provided exemplary
scripts. One should note that both \pkgName{LightGraphs.jl} and \pkgName{PyPlot.jl}
modules can be installed using \julia{Pkg.add} function. Examples in
\file{.ipynb} extensions need \pkgName{GraphPlot.jl} and \pkgName{TikzGraphs.jl}
modules additionally.

%%%%%%%%%%%%%%%%%%%%%%%%%%%%%%%%%%%%%%%%%%%%%%%%%%%%%%%%%%%%%%%%%%%%%%%%%%%%%%%%
\subsection{Provided functions}\label{app:functions}
%%%%%%%%%%%%%%%%%%%%%%%%%%%%%%%%%%%%%%%%%%%%%%%%%%%%%%%%%%%%%%%%%%%%%%%%%%%%%%%%
Below we provide the description of the functionality implemented by \package\
package. The package introduces 4 data types and defines 22 functions. All names
exported from the package can be inspected using \julia{names(QSWalk)} command.
They are grouped into five categories, based on their purpose.

\Julia{} interpreter provides access to the function documentation. In order to 
show description of the function \julia{some_function} one 
should type
\begin{lstlisting}
julia> ?some_function
\end{lstlisting}
One should note that \julia{?} mark disappears and the prompt is changed from
\julia{julia>} to \julia{help?>}. 

%%%%%%%%%%%%%%%%%%%%%%%%%%%%%%%%%%%%%%%%%%%%%%%%%%%%%%%%%%%%%%%%%%%%%%%%%%%%%%%%
% external LaTeX file with the documentation of all functions
%%%%%%%%%%%%%%%%%%%%%%%%%%%%%%%%%%%%%%%%%%%%%%%%%%%%%%%%%%%%%%%%%%%%%%%%%%%%%%%%
%%%%%%%%%%%%%%%%%%%%%%%%%%%%%%%%%%%%%%%%%%%%%%%%%%%%%%%%%%%%%%%%%%%%%%%%%%%%%%%%
\subsubsection{Data types}
%%%%%%%%%%%%%%%%%%%%%%%%%%%%%%%%%%%%%%%%%%%%%%%%%%%%%%%%%%%%%%%%%%%%%%%%%%%%%%%%
Package \package\ takes advantage of the strong typing capabilities of \Julia\
language. Data types defined in the package are used to represented internal
data utilised during the calculations and enable the proper utilization of the
functionality based on sparse matrices.

\begin{itemize}
	\item \julia{SparseDenseMatrix} -- type representing matrices which can be
	dense or sparse.
	\item \julia{SparseDenseVector} -- type representing vectors which can be
	dense or sparse.
\end{itemize}

Types \julia{Vertex} and \julia{VertexSet} are used for the nonmoralizing
evolution. In this case, the standard basis is partitioned for vertices. These
two types are used to store partition description.

\begin{itemize}
	
	\item \julia{Vertex} -- type describing the labels of vectors from the
	canonical basis corresponding to given \julia{Vertex}. Function
	\julia{subspace(Vertex)} returns the list of labels, while \julia{Vertex[i]}
	return a unique label.
	
	\item \julia{VertexSet} -- type consisting of a list of \julia{Vertex}
	objects. It describes the partition of the linear subspace. Objects of this
	type should be constructed with \julia{make_vertex_set} or \julia{nm_lind}
	functions. In order to get a list of the vertices one should use
	\julia{vertices} function, or \julia{vertexset[i]} for a \julia{i}-th
	\julia{Vertex}.
	
\end{itemize}

%%%%%%%%%%%%%%%%%%%%%%%%%%%%%%%%%%%%%%%%%%%%%%%%%%%%%%%%%%%%%%%%%%%%%%%%%%%%%%%%
\subsubsection{Construction of generators}
%%%%%%%%%%%%%%%%%%%%%%%%%%%%%%%%%%%%%%%%%%%%%%%%%%%%%%%%%%%%%%%%%%%%%%%%%%%%%%%%
This group of functions provides the methods for construction of generators of
the dynamical subgroups from the graph representation.

\begin{itemize} 
	
	\item \julia{local_lind(A[, epsilon])} splits the elements of the matrix
	\julia{A} into a collection of sparse matrices with exactly one non-zero
	element. Matrices are added if the absolute value of the nonzero element is
	at least \julia{epsilon}, which defaults to \julia{eps()}.
	
	\item \julia{evolve_generator(H, L[, localH][, omg])} creates the generator
	for the evolvution as in Eq.~\eqref{eq:evolve-generator}, calculated
	according to Eq.~\eqref{eq:GKSL-master-equation},
	\eqref{eq:evolution-smoothing} or
	\eqref{eq:evolution-smoothing-nonmoralizing}, given Hamiltonian \julia{H},
	collection of Lindblad operators \julia{L}, local Hamiltonian \julia{localH}
	and scaling parameter \julia{omg}. Parameters \julia{localH} and \julia{omg}
	are optional. If \julia{omg} is not provided, $(1-\omega)$ and $\omega$ parts
	in Eq.~\eqref{eq:evolution-smoothing-nonmoralizing} are set to 1.
	
\end{itemize}

%%%%%%%%%%%%%%%%%%%%%%%%%%%%%%%%%%%%%%%%%%%%%%%%%%%%%%%%%%%%%%%%%%%%%%%%%%%%%%%%
\subsubsection{Evolution}
%%%%%%%%%%%%%%%%%%%%%%%%%%%%%%%%%%%%%%%%%%%%%%%%%%%%%%%%%%%%%%%%%%%%%%%%%%%%%%%%

Functions in this group provide the interface for simulating quantum stochastic
walks in the local and the global regimes. The functions use different approach
if the evolution generator is sparse or dense. For a dense matrix its
exponentiation is calculated, which is efficient for small matrices. For a
sparse matrix \julia{expmv} provided by \package{Expokit} package is used
\cite{expokit}. Note that sparsity is checked by the type of the evolution
generator.

\begin{itemize}

    \item \julia{evolve(evo_gen, init_state, time)} is the simplest case where
    the function accepts evolution generator \julia{evo_gen}, see
    Eq.~\eqref{eq:evolve-generator}, \julia{init_state} describing the initial
    state of the evolution, and \julia{time} specifying the time of the
    evolution. Argument \julia{time} has to be non-negative.

    \item \julia{evolve(evo_gen, init_state, tpoints)} is similar to the
    previous one, but a list of points of time \julia{tpoints} is given. In this
    case, a list of the resulting states is returned. This results in speedup
    where \julia{evo_gen} is a dense matrix.
\end{itemize}

If the evolution is applied to several initial states and the evolution
generator operator is dense, it is more efficient to calculate the superoperator
first, and next to apply it to the initial states. For this scenario the package
provides two functions.

\begin{itemize}
    \item \julia{evolve_operator(evo_gen, time)} returns an exponent of
    \julia{time}$\times$\julia{evo_gen} called superoperator. The function works
    for dense matrices only.

    \item \julia{evolve(evo_super, init_state)} returns a state based on
    superoperator generated by \julia{evolve_operator}. The function works for
    dense matrices only.
\end{itemize}

It is important to note that the user must provide input arguments fulfilling
the appropriate conditions. For the procedure \julia{evolve} to work correctly,
\julia{evo_gen} should be generated by \julia{evolve_generator} function and
\julia{init_state} should be a proper density matrix.

%%%%%%%%%%%%%%%%%%%%%%%%%%%%%%%%%%%%%%%%%%%%%%%%%%%%%%%%%%%%%%%%%%%%%%%%%%%%%%%%
\subsubsection{Demoralization}
%%%%%%%%%%%%%%%%%%%%%%%%%%%%%%%%%%%%%%%%%%%%%%%%%%%%%%%%%%%%%%%%%%%%%%%%%%%%%%%%

The functions in this group provides functionality required to construct
nonmoralizing evolution on directed graphs. This is necessary to reproduce the
structure of directed graphs using quantum stochastic walks. In this case, one
needs to provide an extended initial state, perform the evolution using a
special Hamiltonian, and perform the measurement which interprets the enlarged
space (see. Fig.~\ref{fig:work-flow}).

\begin{itemize} 
    \item \julia{default_nm_loc_ham(size)} returns a default part of a local
    Hamiltonian of size \julia{size}$\times$\julia{size} for vertex subspace of
    given order. The Hamiltonian is sparse with nonzero elements on the first
    upper diagonal equal to \julia{1im} and the lower diagonal equal to
    \julia{-1im}.

    \item \julia{nm_loc_ham(vertexset, hams)} returns a Hamiltonian acting
    locally on each vertex from \julia{vertexset} partition. Optional argument
    \julia{hams} is a dictionary which, for a given dimension of vertex linear
    subspace, yields a hermitian operator. It can be \julia{Dict\{Int,
    SparseDenseMatrix\}}, which returns the matrix by the indegree, or
    \julia{Dict\{Vertex, SparseDenseMatrix\}} which, for different vertices, may
    return different matrices.

    \item \julia{nm_lind(A[, linds][, epsilon])])} returns a single Lindbladian
    operator and a vertex set describing how vertices are bound to subspaces.
    The operator is constructed according to the correction scheme presented in
    \cite{domino2017superdiffusive}. Parameter \julia{A} is a square matrix,
    describing the connection between the canonical subspaces in a similar
    manner as the adjacency matrix. Parameter \julia{epsilon}, with the
    default value \julia{eps()}, determines the relevant values by
    \julia{abs(A[i, j])>=epsilon} formula. List \julia{linds} describes the
    elementary matrices. It can be \julia{Dict\{Int, SparseDenseMatrix\}}, which
    returns the matrix by the indegree, or \julia{Dict\{Vertex,
    SparseDenseMatrix\}} which, for different vertices, may return different
    matrices. The matrix should have orthogonal columns and be of the size
    outdegree of the vertex. As the default, the function uses Fourier matrices
    (see. Sec.~\ref{sec:doc-dirac}).
    
    \item \julia{nm_glob_ham(A[, hams][, epsilon])} returns a global Hamiltonian
    for the moralization procedure. Matrix \julia{A} should be a symmetric
    matrix, for which one aims to construct the nonmoralizing dynamics. Here,
    \julia{hams} is an optional argument which is a dictionary with keys of type
    \julia{Tuple\{Int, Int\}} or \julia{Tuple\{Vertex, Vertex\}}. The first one
    collects the submatrices according to their shape, while the second one
    collects them according to each pair of vertices. As the default, all-one
    submatrices are chosen. The last argument states that only the elements for
    which \julia{abs(A[i, j])>epsilon} are considered.
    
    \item \julia{nm_measurement(probability, vertexset)} returns the joint
    probability of \julia{probability}, which is real-valued probability vector
    according to partition \julia{vertexset}.

    \item \julia{nm_measurement(state, vertexset)} returns the joint probability
    of canonical measurement of density matrix \julia{state}, according to
    partition \julia{vertexset}.
    
    \item \julia{nm_init(init_vertices, vertexset)} returns the initial state in
    the case of the nonmoralizing evolution. The result is a block diagonal
    matrix, where each block corresponds to vertex from \julia{vertexset}. If
    the first argument is of type \julia{Vector\{Vertex\}}, then default block
    matrix \julia{eye} is used. Note that in this case the initial density state
    is normalized in a uniform distribution way, \ie different vertices have the
    same probability.
    
    \julia{nm_init(init_states, vertexset)} -- similar to the function above,
    the first argument is of type \julia{Dict\{Vertex, SparseDenseMatrix\}}. For
    each given vertex a block from dictionary is used, otherwise zero matrix is
    chosen. Each matrix from dictionary should be nonnegative and sum of all
    traces should equal one. The keys of \julia{init_vertices} should be a
    subset of \julia{vertices(vertexset)}. Note that the matrix from \julia{init_states}
    corresponding to vertex \julia{v} should be of size
    \julia{length(v)}$\times$\julia{length(v)}.

\end{itemize}

%%%%%%%%%%%%%%%%%%%%%%%%%%%%%%%%%%%%%%%%%%%%%%%%%%%%%%%%%%%%%%%%%%%%%%%%%%%%%%%%
\subsubsection{Dirac notation and matrix utilities}\label{sec:doc-dirac}
%%%%%%%%%%%%%%%%%%%%%%%%%%%%%%%%%%%%%%%%%%%%%%%%%%%%%%%%%%%%%%%%%%%%%%%%%%%%%%%%
The last category of functions contains procedures used to manipulate matrices
and vectors~\cite{miszczak2011singular}. These functions can be used
independently from the rest of the package. Note that \Julia package indexes
list from 1, and similarly all relevant functions require a positive
\julia{index}.

\begin{itemize} 
	
	\item \julia{ket(index, size)} returns \julia{index}-th column vector from
	standard basis in the \julia{size}-dimensional vector space.
	
	\item \julia{bra(index, size)} returns \julia{index}-th row vector from
	standard basis in the \julia{size}-dimensional vector space.
	
	\item \julia{ketbra(irow, icol, size)} return a matrix acting on
	\julia{size}-dimensional vector space. The matrix consists of a single
	non-zero element equal to one, located at (\julia{irow}, \julia{icol}).
	
	\item \julia{proj(index, size)} returns a projector onto \julia{index}-th
	base vector in \julia{size}-dimensional vector space. This is equivalent to
	\julia{ketbra(index, index, size)}.
	
	\item \julia{proj(vector)} returns a projector onto the subspace spanned by
	vector \julia{vector}.
	
	\item \julia{res(mtx)} returns the vectorization of the matrix \julia{mtx}
	in the row order. This is equivalent to \julia{Base.vec(transpose(mtx))}.
	
	\item \julia{unres(vector)} -- inverse of \julia{res} function. 
	
	\item \julia{fourier_matrix(dim)} returns Fourier matrix of size
	\julia{dim}$\times$\julia{dim}.
	
\end{itemize}

%%%%%%%%%%%%%%%%%%%%%%%%%%%%%%%%%%%%%%%%%%%%%%%%%%%%%%%%%%%%%%%%%%%%%%%%%%%%%%%%

%%%%%%%%%%%%%%%%%%%%%%%%%%%%%%%%%%%%%%%%%%%%%%%%%%%%%%%%%%%%%%%%%%%%%%%%%%%%%%%%
\subsection{Typical work-flow}
%%%%%%%%%%%%%%%%%%%%%%%%%%%%%%%%%%%%%%%%%%%%%%%%%%%%%%%%%%%%%%%%%%%%%%%%%%%%%%%%
\begin{figure}[hp]   
\begin{tikzpicture}[scale=0.5,node distance=.9cm, start chain=going below,]
\tikzset{
	>=stealth',
	selection/.style={ ellipse, draw=black, thick,
		text width=8em,text centered,
		on chain},
	qswalkFunctions/.style={ rectangle, rounded corners, draw=black, thick,
		text width=11em, minimum height=3em, text centered,
		on chain},               
	outerFunctions/.style={ rectangle, draw=black, thick,
		text width=11em, minimum height=3em, text centered,
		on chain},               
	decision/.style = { diamond, draw=black, thick,
		text width=4em, text centered,
		on chain},
	line/.style={draw, thick, <-},
	every join/.style={->, thick, shorten >= 1pt},  
}

%  \node[selection, on chain=going right] (params) {Simulation parameters\\(dimension, init state)};
%  \node[selection, on chain=going right, join = with params] {Graph\\(adjacency matrix)};
  \node[outerFunctions, on chain=going right] {Graph\\(adjacency matrix)};
  \node[decision, join] (regime) {Regime selection};
  \node[selection] (global) [below = of regime, join = with regime] {Global interaction};
  \node[selection] (local) [left = of global, join = with regime] {Local interaction};
  \node[selection] (nonmoral) [right = of global, join = with regime] {Non-moralizing global interaction};
  
  \node[outerFunctions] (local-ham) [below = of local, join = with local] {H = adjacency matrix};
  \node[qswalkFunctions] (local-lin) [below = of local-ham, join = with local-ham] {$\mathcal{L} =$ \julia{local_lind}};
	\node[outerFunctions] (local-init) [below = of local-lin, join = with local-lin] {initial state};
  
  \node[outerFunctions] (global-ham) [below = of global, join = with global] {H = adjacency matrix };
   \node[outerFunctions] (global-lin) [below = of global-ham, join = with global-ham] { $\mathcal{L} = \{\textrm{adjacency\ matrix} \}$};
   	\node[outerFunctions] (global-init) [below = of global-lin, join = with global-lin] {initial state};
   
  \node[qswalkFunctions] (nonmoral-ham) [below = of nonmoral, join = with nonmoral,yshift=.5cm] {H = \julia{nm_glob_ham} };
\node[qswalkFunctions] (nonmoral-lin) [below = of nonmoral-ham, join = with nonmoral-ham,yshift=.5cm] {L = \{\julia{nm_lind}\}};
  \node[qswalkFunctions] (nonmoral-loc) [below = of nonmoral-lin, join = with nonmoral-lin,yshift=.5cm] {$H_{\mathrm{loc}}$ = \julia{nm_loc_ham} };
  \node[qswalkFunctions] (nonmoral-init) [below = of nonmoral-loc, join = with nonmoral-loc, yshift=.5cm] {initial state \julia{nm_init}};
 
  \node[qswalkFunctions] (evolve-op)  [below = of global-init, join = with global-init, join = with local-init, join = with nonmoral-init] {\julia{evolve_generator}};
  \node[qswalkFunctions] (evolve) [join = with evolve-op] {\julia{evolve}};
  
  \node[outerFunctions] (measure) at ($(local-lin)!0.5!(global-lin)$) [join = with evolve, yshift= -6.8cm]{measure with \julia{diag}};  
  \node[qswalkFunctions] (measure) [below = of nonmoral-loc, join = with evolve, yshift= -5.14cm]{measure with \julia{nm_measurement}};  
%  \node[selection] (nonmoral-init) [below = of nonmoral, join = with nonmoral] 
%  {Initilize extended initial state};
%  \node[selection] (nonmoral-evolv) [below = of nonmoral-init, join = with nonmoral-init] {Evolve on the extended state};
%  \node[selection] (nonmoral-measure) [below = of nonmoral-evolv, join = with nonmoral-evolv] {Measure on the equivalence classes};
\end{tikzpicture}
\caption{Typical steps required during the preparation of the simulation
utilizing \package\ package. The rounded rectangles refer to the functions
implemented in \package, while the standard rectangles require user input which
should be provided by some external functions, \eg for \julia{adjacency_matrix}
from \pkgName{LightGraphs.jl}.} \label{fig:work-flow}
\end{figure}
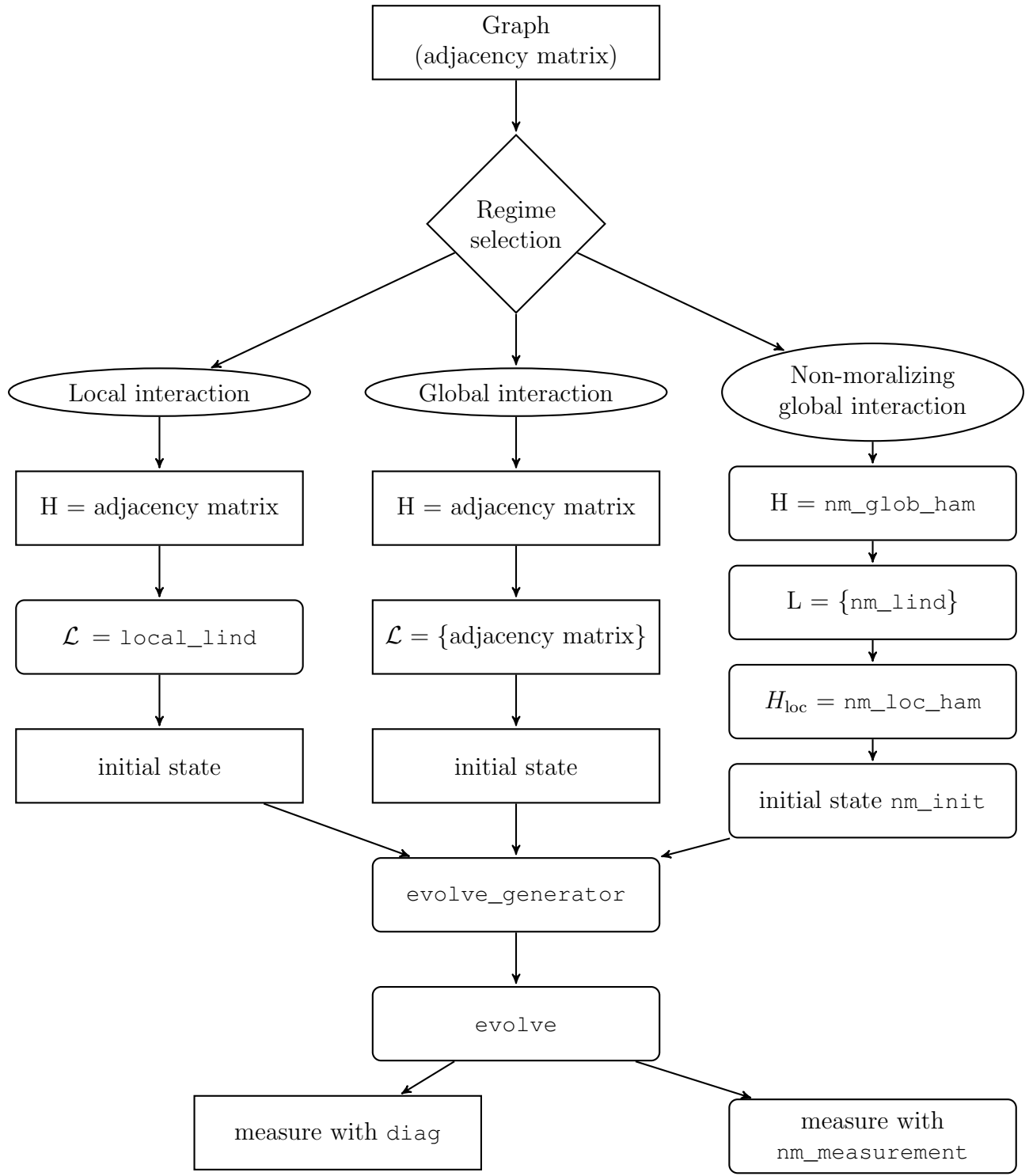

Typical work-flow is presented in Fig.~\ref{fig:work-flow}. For local and global
regimes, all operators should be generated by the user. This includes the
Hamiltonian, Lindblad operators and the initial state. In the local regime
case, the \package package provides matrix splitting function, which generates
local operations of the form $\ketbra{v}{w}$. 

Nonmoralizing interaction requires additional steps. We start with the global
regime Hamiltonian and a single Lindblad operator. Based on them, we construct
nonmoralizing operators through \julia{nm_glob_ham} and \julia{nm_lind}
functions. The user is encouraged to construct the local Hamiltonian by
\julia{nm_loc_ham} function. Function \julia{nm_init} should be used to
construct block-diagonally mixed state, where each block corresponds to a
vertex.

Independently of the chosen regime, the Hamiltonian, the Lindblad operators and
optional local Hamiltonian should be passed to \julia{evolve_generator}, which
constructs evolution generator according to the appropriate formula, and then
pass it to \julia{evolve}. At this point, we have a density matrix. In order to
measure it, for local and global regimes \julia{diag} function refers to canonic
measurement. For nonmoralizing global interaction we provide a separate function
\julia{nm_measurement}. 

%%%%%%%%%%%%%%%%%%%%%%%%%%%%%%%%%%%%%%%%%%%%%%%%%%%%%%%%%%%%%%%%%%%%%%%%%%%%%%%%
\section{Usage examples} 
%%%%%%%%%%%%%%%%%%%%%%%%%%%%%%%%%%%%%%%%%%%%%%%%%%%%%%%%%%%%%%%%%%%%%%%%%%%%%%%%

In this section we provide some basic examples, which present functionalities of
\package module. In particular, we demonstrate the examples with all described
evolution regimes for quantum stochastic walk, namely local environment
interaction, global environment interaction, and nonmoralizing global
environment interaction. The examples are based on the results presented in
\cite{domino2016properties,glos2017limit,liu_continuous-time_2016,domino2017superdiffusive,bringuier_central_2016}.

Code snippets in this section can be found in the form of full examples in the
\texttt{examples} subdirectory, contained in the package distribution. For
convenience we provide the examples in the form of \Julia\ scripts (files with
\file{.jl} extension), as well as \Jupyter\ notebooks (corresponding files with
\file{.ipynb} extension). Some of the examples require \pkgName{LightGraphs.jl} and
\pkgName{PyPlot.jl} modules. We use \julia{#} (hash), denoting a comment in \Julia,
to provide the result of the command.
%Standard \Julia commands are typeset using bold font.

%%%%%%%%%%%%%%%%%%%%%%%%%%%%%%%%%%%%%%%%%%%%%%%%%%%%%%%%%%%%%%%%%%%%%%%%%%%%%%%%
\subsection{Propagation on the line segment}
%%%%%%%%%%%%%%%%%%%%%%%%%%%%%%%%%%%%%%%%%%%%%%%%%%%%%%%%%%%%%%%%%%%%%%%%%%%%%%%%

\begin{figure}[ht!]\centering
	\subfigure[Global interaction regime
	]{\includegraphics[width=0.45\textwidth]{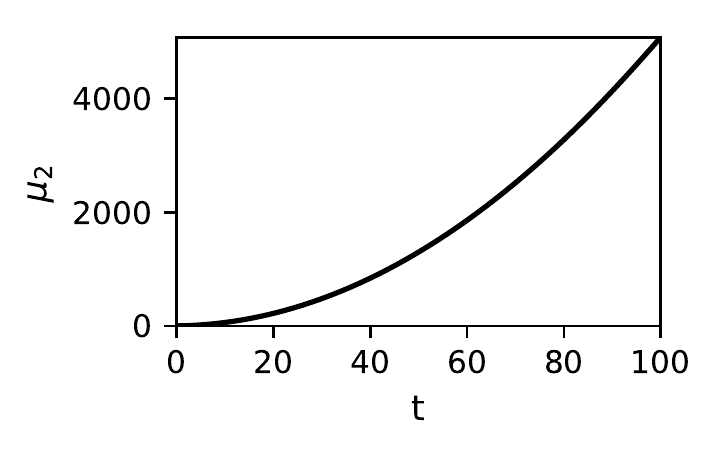}\label{fig:secondmomentglobal}}
	\subfigure[Local interaction regime	
	]{\includegraphics[width=0.45\textwidth]{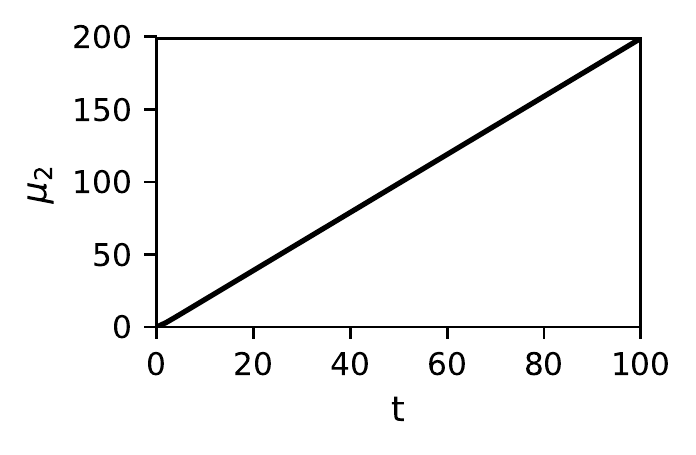}\label{fig:secondmomentlocal}}
  \caption{Second central moment as the function of time for the global
  \subref{fig:secondmomentglobal} and the local \subref{fig:secondmomentlocal}
  interaction regime. The value of the second moment  grows ballistically for
  the global interaction case.} \label{fig:propagation}
\end{figure}

The first example demonstrates the evolution on the line segment using 
different regimes. Code snippets from this section can be found in the 
\file{ex01_path_propagation.jl} file.
The plots resulting from the above examples are presented in
Fig.~\ref{fig:propagation}. One can note that the second moment for the global
interaction case grows ballistically.

We would like to recover the results from \cite{domino2016properties}
corresponding to the analysis of the second moment of the walker position. In
the first step we prepare the essential data.
\begin{lstlisting}
dim = 251
w = 0.5
timepoints = collect(0:2:100)
adjacency = adjacency_matrix(PathGraph(dim))
\end{lstlisting}
We consider both global and local interaction cases. Hence we need to prepare 
the evolution generators corresponding to both types of evolution.
\begin{lstlisting}
lind_local = local_lind(adjacency)
midpoint = ceil(Int, dim/2)
op_global = evolve_generator(adjacency, [adjacency], w)
op_local = evolve_generator(adjacency, lind_local, w)
\end{lstlisting}
Note that \julia{evolve_generator} accepts a list of matrices as the second 
argument and one needs to put extra square bracket for \julia{adjacency} 
parameter in \julia{op_global} case. Next, 
using the created operators, we simulate the evolution using \julia{evolve} 
function.
\begin{lstlisting}
rho_global = evolve(op_global, proj(midpoint, dim), timepoints)
rho_local = evolve(op_local, proj(midpoint, dim), timepoints)
\end{lstlisting}
Note that in both cases the second argument refers to the initial state of the
evolution, which is at the middle-point of the path graph. Finally, we can
calculate the second central moment of the distribution for the standard
measurement of the position as
\begin{lstlisting}
secmoment_global = Float64[]
secmoment_local = Float64[]
positions = (collect(1:dim)-midpoint)
for i=1:length(timepoints)
   push!(secmoment_global, sum(positions.^2 .* diag(rhoglobal[i])))
   push!(secmoment_local, sum(positions.^2 .* diag(rholocal[i])))
end
\end{lstlisting}
and prepare the plots
\begin{lstlisting}
plot(timepoints, secmoment_global)
plot(timepoints, secmoment_local)
\end{lstlisting}

%%%%%%%%%%%%%%%%%%%%%%%%%%%%%%%%%%%%%%%%%%%%%%%%%%%%%%%%%%%%%%%%%%%%%%%%%%%%%%%%
\subsection{Convergence on random graphs}
%%%%%%%%%%%%%%%%%%%%%%%%%%%%%%%%%%%%%%%%%%%%%%%%%%%%%%%%%%%%%%%%%%%%%%%%%%%%%%%%

In this example we demonstrate the use of \julia{evolve_operator} function for
the purpose of analysing the convergence of the quantum stochastic walks. In
this situation we are interested in the behaviour of the process, represented by
the evolve generator, on different initial states. The code described here can be
found in supplementary materials in \file{ex02_convergence.jl}.

Quantum stochastic walks have convergence properties different from typical
quantum walk models. In particular, for the local environment interaction
regime, for arbitrary strongly connected directed graph the evolution is
relaxing, \ie there exists a unique density matrix which is the limiting state
for the evolution for an arbitrary initial state
\cite{liu_continuous-time_2016,glos2017limit}. 

Let us consider a random directed graph generated according to Erd\H{o}s-R\'enyi
model.
\begin{lstlisting}
dim = 10
digraph = erdos_renyi(dim, 0.5, is_directed=true)
graph = Graph(digraph)
adj_digraph = full(adjacency_matrix(digraph, :in))
adj_graph = full(adjacency_matrix(graph))
time = 100.
\end{lstlisting}
The Lindbladian and the subgroup generator corresponding to the above graph are 
constructed using \julia{local_lind} and \julia{evolve_generator} 
functions, respectively
\begin{lstlisting}
lind = local_lind(adj_digraph)
evo_gen = evolve_generator(adj_graph, lind)
\end{lstlisting}

In this case, due to the small size of the analysed graph, it is better to
choose the dense matrix type instead of the sparse one. In order to show the
uniqueness of stationary state, it is enough to compute the dimensionality of
the null-space of global operator matrix.
\begin{lstlisting}
println(count(x->abs(x)<1e-5, eigvals(evo_gen)))
# 1
\end{lstlisting}

One can see that dimensionality equals one, and hence there is a unique
stationary state. The existence of at least one stationary state is guaranteed
by quantum Perron-Frobenius theorem~\cite{albeverio1978frobenius}.

To check that this is indeed the case we can choose different initial states to 
show that the stationary state is unique.
\begin{lstlisting}
rhoinit1 = proj(1, dim)
rhoinit2 = proj(3, dim)
rhoinit3 = eye(dim)/dim
\end{lstlisting}
Since we simulate a single evolution for all of the above states, for dense
matrices it is better to compute the exponential once, and then use it for all
of the states.
\begin{lstlisting}
U = evolve_operator(evo_gen, time)
rho1 = evolve(U, rhoinit1)
rho2 = evolve(U, rhoinit2)
rho3 = evolve(U, rhoinit3)

println(norm(rho1-rho2))
println(norm(rho2-rho3))
# 7.001526329112005e-17
# 4.1919257307409924e-17
\end{lstlisting}

One should note that, since the graph is chosen randomly from Erd\H{o}s-R\'enyi 
distribution, the resulting number will be different for each execution of the 
above code. 

%%%%%%%%%%%%%%%%%%%%%%%%%%%%%%%%%%%%%%%%%%%%%%%%%%%%%%%%%%%%%%%%%%%%%%%%%%%%%%%%
\subsection{Spontaneous moralization}
%%%%%%%%%%%%%%%%%%%%%%%%%%%%%%%%%%%%%%%%%%%%%%%%%%%%%%%%%%%%%%%%%%%%%%%%%%%%%%%%
In this section we present the basics of the spontaneous moralization
\cite{domino2017superdiffusive}. The code described here can be found in
supplementary materials in \file{ex03_moralization_simple.jl}. Here we plan to
analyse the graph presented in Fig.~\ref{fig:moralization-example}. Let us
analyse the case where the global Lindblad operator equal to adjacency matrix is
the only operator used. Again we start with preparing the global operator
\begin{lstlisting}
adjacency = [0 0 0;
             0 0 0;
             1 1 0]
opmoral = evolve_generator(zero(adjacency), [adjacency])
time = 100.
\end{lstlisting}
In the evolution we start in the top-left vertex. One could expect that for
large time the state will converge to sink (bottom) vertex. However running the
following commands shows, that it is not the case.
\begin{lstlisting}
rho = evolve(opmoral, proj(1,3), time)
println(diag(rho))
# Complex{Float64}[0.25+0.0im, 0.25+0.0im, 0.5+0.0im]
\end{lstlisting}
We have a non-zero probability of measuring top vertices. Whats more, the
amplitude transited from vertex top-left to top-right, although there is no path
between them. 

The key is to choose nonmoralizing procedure. We need to create new operators, called nonmoralizing Lindbladian operator and local Hamiltonian.
\begin{lstlisting}
lnonmoral, vset = nm_lind(adjacency)
hlocal = nm_loc_ham(vset)
opnonmoral = evolve_generator(zero(lnonmoral), [lnonmoral], hlocal)
\end{lstlisting}
Note that new parameter \julia{vset} appears. The nonmoralizing procedure needs
description how the linear space is divided into subspaces corresponding to
vertices. Usually, \julia{vset} should be treated as the parametrization of the
evolution, which should be passed for all of the functions without change. It
can be generated by \julia{nm_lind}, as in the example above, or simply by
\julia{make_vertex_set}.

Now, the operators have bigger size than the original ones. The size can be
calculated with \julia{vertexsetsize} function.
\begin{lstlisting}
println(vertexsetsize(vset))
# 4
println(vset)
# QSWalk.VertexSet(QSWalk.Vertex[QSWalk.Vertex([1]), QSWalk.Vertex([2]),
# QSWalk.Vertex([3, 4])])
\end{lstlisting} 
Here \julia{vset} describes the partition of linear space, in this example
$\spann(\ket 1)$ corresponds to the first vertex, $\spann(\ket 2)$ to the second
vertex and $\spann(\ket 3, \ket 4)$ to the third vertex. While it is possible to
write own density states, for states being block diagonal, where each block
corresponds to a different vertex, simpler function can be chosen.
\begin{lstlisting}
rho0 = nm_init(vset[[1]], vset)
\end{lstlisting}
Note that as the first argument we choose a list of vertices. The state is an
identity matrix on a subspace corresponding to \julia{rho0}. In this case after
the evolution we obtain
\begin{lstlisting}
rho = evolve(opnonmoral, rhoinit, time)
println(nm_measurement(rho, vset))
# [5.02465e-33, -5.74849e-33, 1.0]
\end{lstlisting}
Now we have the expected result, \ie{} all of the amplitudes are transfered into
the sink (bottom vertex).

%%%%%%%%%%%%%%%%%%%%%%%%%%%%%%%%%%%%%%%%%%%%%%%%%%%%%%%%%%%%%%%%%%%%%%%%%%%%%%%%
\subsection{Spontaneous moralization on path graph}
%%%%%%%%%%%%%%%%%%%%%%%%%%%%%%%%%%%%%%%%%%%%%%%%%%%%%%%%%%%%%%%%%%%%%%%%%%%%%%%%
In this section we recover the results presented in
\cite{domino2017superdiffusive}. The code described here can be found in
supplementary materials in \file{ex04_moralization_path.jl}. Demoralizing
procedure results in non-symmetric evolution. Let us again consider the code
below.
\begin{lstlisting}
dim = 101 #odd for unique middle point
w = 0.5
time = 40.
adjacency = adjacency_matrix(PathGraph(dim))

midpoint = ceil(Int, dim/2)
lind, vset = nm_lind(adjacency)
hglobal = nm_glob_ham(adjacency)
hlocal = nm_loc_ham(vset)
opnonsymmetric = evolve_generator(hglobal, [lind], hlocal, w)
rhoinit = nm_init(vset[[midpoint]], vset)

rho_nonsymmetric = evolve(opnonsymmetric, rhoinit, time)
\end{lstlisting}
For the canonical measurement, the distribution should be symmetric with respect
to \julia{midpoint}, and hence the commands
\begin{lstlisting}
positions = (collect(1:dim)-midpoint)
measurement_nonsymmetric = nm_measurement(rho_nonsymmetric, vset)
println(sum(positions .* measurement_nonsymmetric))
\end{lstlisting}
should print 0. However, the output is approximately $-0.5698$. Further analysis 
would show that the situation happens even after removing \julia{Hlocal} or 
\julia{H} operators. 

The key is to add another Lindbladian operator, which make the evolution
symmetric again. This can be done by proper adjustment of \julia{nm_lind}.
\begin{lstlisting}
linddescription1 = Dict(1 => ones(1,1), 2 => [1 1; 1 -1])
linddescription2 = Dict(1 => ones(1,1), 2 => [1 1; -1 1])
lind1, vset = nm_lind(adjacency, linddescription1)
lind2, vset = nm_lind(adjacency, linddescription2)
hglobal = nm_glob_ham(adjacency)
hlocal = nm_loc_ham(vset)
opsymmetric = evolve_generator(hglobal, [lind1, lind2], hlocal, w)
rhoinit = nm_init(vset[[midpoint]], vset)
\end{lstlisting}
If we leave the rest of the code unchanged, then we will receive
approximately \julia{-6.759E-9} as an output, which can be treated as numerical zero. Note
that for \julia{nm_lind} we can either make a dictionary for all degrees, or for
all vertices. Similarly, \julia{nm_glob_ham} can be generalized for all pairs of
degrees or all pairs of vertices.

%%%%%%%%%%%%%%%%%%%%%%%%%%%%%%%%%%%%%%%%%%%%%%%%%%%%%%%%%%%%%%%%%%%%%%%%%%%%%%%%
\subsection{Performance analysis} \label{sec:performance}
%%%%%%%%%%%%%%%%%%%%%%%%%%%%%%%%%%%%%%%%%%%%%%%%%%%%%%%%%%%%%%%%%%%%%%%%%%%%%%%%

At the moment of writing, the only package developed for simulating the
evolution of quantum stochastic walks is \pkgName{QSWalk.m} package for \WMMA.
Hence we compared the performance of computing non-moralizing evolution on path
graph for various orders.

\begin{figure}[ht!]
    \centering \includegraphics[width=0.65\columnwidth]{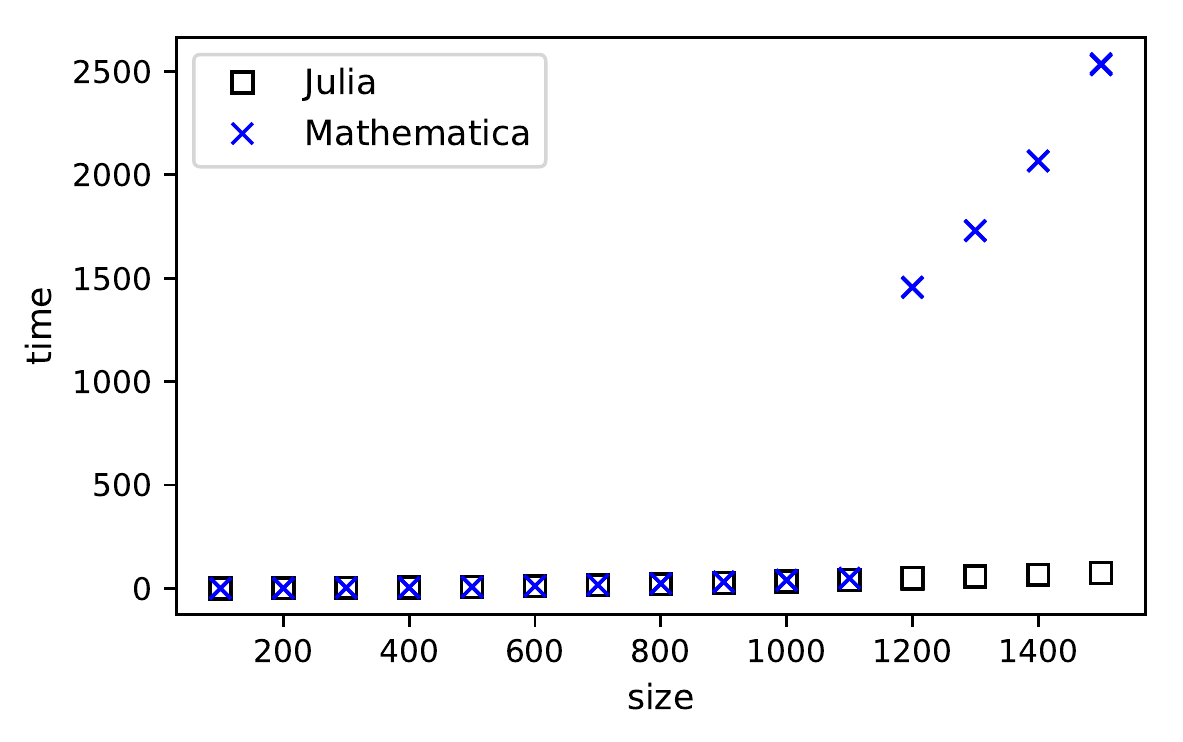}
\caption{Comparison of time required to simulated quantum stochastic walk on
	the line in Julia with \package\ and in \MMA, using the package described in
	\cite{falloon2017qswalk}. Parameter \julia{size} denotes the size of the
	graph. The plot was generated using \MMA 11.2 and \Julia{} 0.6 on Linux
	x86\_64 on Intel Core i7-7700 processor.}\label{fig:mathematica-vs-julia}
\end{figure}

Results presented in Fig.~\ref{fig:mathematica-vs-julia} demonstrate that for
small enough graphs \Julia and \MMA compute the result in similar time. However,
after some order threshold we can observe sudden drop of efficiency of \MMA. Our
in-depth analysis showed that the main reason is the change of \julia{MatrixExp}
function behavior. Unfortunately, \MMA is not an open source programming
language, hence we cannot find any explanation of the phenomena. Most probably
then inefficiency calculation of the results from internal aspects of \MMA
implementation. The calculations were executed on three different architectures
and all of them yield efficiency drop for \pkgName{QSWalk.m} package for \WMMA
for the same order of the input matrix.

%%%%%%%%%%%%%%%%%%%%%%%%%%%%%%%%%%%%%%%%%%%%%%%%%%%%%%%%%%%%%%%%%%%%%%%%%%%%%%%%
\section{Concluding remarks and future work}
%%%%%%%%%%%%%%%%%%%%%%%%%%%%%%%%%%%%%%%%%%%%%%%%%%%%%%%%%%%%%%%%%%%%%%%%%%%%%%%%

In this \docName we have provided the description of \package module developed
for the purpose of studying quantum stochastic walks. The packages is focused on
the graph-based evolution and provides many functions typically used for quantum
walks on graphs. We have utilized \Julia programming language and the package
demonstrates an advantage over the existing software in terms of the
performance. The nonmoralizing evolution first implemented in the package
enables the study of quantum procedures developed using quantum stochastic walks
on arbitrary directed graphs. As such, the package can be used to gain more
insight into the properties of quantum evolution on this type of graphs. While
the name of the presented package refers to Quantum Stochastic Walk, it is not
limited to this type of evolution only. Many of the the presented functions can
be used for arbitrary quantum evolution with constant $H$ and $\LL$.

%The efficiency of the presented package depends on the times required to
%calculate the matrix exponentiation, which is implemented by \julia{expmv}
%function. This function is provided by \pkgName{Expokit} package and is beyond
%the scope of the presented work. However, we plan to improve the user-interface
%delivered by the \package. Currently we are working on the general quantum walk
%interface \pkgName{QSpatialSearch.jl} designed for the analysis of various
%models of quantum walks. 
%
%We plan to adjust current package version to new syntax, and by this `import'
%functionalities given by the package. Before that we plan to register the
%package in official \Julia repository. As \Julia is still an evolving language,
%we plan to adjust the package to newer version of Julia, including 1.0 version.

%%%%%%%%%%%%%%%%%%%%%%%%%%%%%%%%%%%%%%%%%%%%%%%%%%%%%%%%%%%%%%%%%%%%%%%%%%%%%%%%
\paragraph{Acknowledgements}
%%%%%%%%%%%%%%%%%%%%%%%%%%%%%%%%%%%%%%%%%%%%%%%%%%%%%%%%%%%%%%%%%%%%%%%%%%%%%%%%
This work has been supported by the Polish Ministry of Science and Higher
Education under project number IP 2014 031073. Authors would like to thank Piotr
Gawron for valuable remarks concerning the usage of type system in \Julia.

%%%%%%%%%%%%%%%%%%%%%%%%%%%%%%%%%%%%%%%%%%%%%%%%%%%%%%%%%%%%%%%%%%%%%%%%%%%%%%%%
\bibliographystyle{apsrev4-1}
\bibliography{qsw-julia}

\begin{thebibliography}{28}%
\makeatletter
\providecommand \@ifxundefined [1]{%
 \@ifx{#1\undefined}
}%
\providecommand \@ifnum [1]{%
 \ifnum #1\expandafter \@firstoftwo
 \else \expandafter \@secondoftwo
 \fi
}%
\providecommand \@ifx [1]{%
 \ifx #1\expandafter \@firstoftwo
 \else \expandafter \@secondoftwo
 \fi
}%
\providecommand \natexlab [1]{#1}%
\providecommand \enquote  [1]{``#1''}%
\providecommand \bibnamefont  [1]{#1}%
\providecommand \bibfnamefont [1]{#1}%
\providecommand \citenamefont [1]{#1}%
\providecommand \href@noop [0]{\@secondoftwo}%
\providecommand \href [0]{\begingroup \@sanitize@url \@href}%
\providecommand \@href[1]{\@@startlink{#1}\@@href}%
\providecommand \@@href[1]{\endgroup#1\@@endlink}%
\providecommand \@sanitize@url [0]{\catcode `\\12\catcode `\$12\catcode
  `\&12\catcode `\#12\catcode `\^12\catcode `\_12\catcode `\%12\relax}%
\providecommand \@@startlink[1]{}%
\providecommand \@@endlink[0]{}%
\providecommand \url  [0]{\begingroup\@sanitize@url \@url }%
\providecommand \@url [1]{\endgroup\@href {#1}{\urlprefix }}%
\providecommand \urlprefix  [0]{URL }%
\providecommand \Eprint [0]{\href }%
\providecommand \doibase [0]{http://dx.doi.org/}%
\providecommand \selectlanguage [0]{\@gobble}%
\providecommand \bibinfo  [0]{\@secondoftwo}%
\providecommand \bibfield  [0]{\@secondoftwo}%
\providecommand \translation [1]{[#1]}%
\providecommand \BibitemOpen [0]{}%
\providecommand \bibitemStop [0]{}%
\providecommand \bibitemNoStop [0]{.\EOS\space}%
\providecommand \EOS [0]{\spacefactor3000\relax}%
\providecommand \BibitemShut  [1]{\csname bibitem#1\endcsname}%
\let\auto@bib@innerbib\@empty
%</preamble>
\bibitem [{jul(2018)}]{julia-docs}%
  \BibitemOpen
  \href@noop {} {\enquote {\bibinfo {title} {Julia documentation},}\ }
  (\bibinfo {year} {2018}),\ \bibinfo {note}
  {\url{https://docs.julialang.org/en/latest/} (Accessed on
  18/12/2017)}\BibitemShut {NoStop}%
\bibitem [{\citenamefont {Bezanson}\ \emph {et~al.}(2017)\citenamefont
  {Bezanson}, \citenamefont {Edelman}, \citenamefont {Karpinski},\ and\
  \citenamefont {Shah}}]{bezanson2017julia}%
  \BibitemOpen
  \bibfield  {author} {\bibinfo {author} {\bibfnamefont {J.}~\bibnamefont
  {Bezanson}}, \bibinfo {author} {\bibfnamefont {A.}~\bibnamefont {Edelman}},
  \bibinfo {author} {\bibfnamefont {S.}~\bibnamefont {Karpinski}}, \ and\
  \bibinfo {author} {\bibfnamefont {V.~B.}\ \bibnamefont {Shah}},\ }\href
  {\doibase 10.1137/141000671} {\bibfield  {journal} {\bibinfo  {journal} {SIAM
  Review}\ }\textbf {\bibinfo {volume} {59}},\ \bibinfo {pages} {65} (\bibinfo
  {year} {2017})}\BibitemShut {NoStop}%
\bibitem [{\citenamefont {Domino}\ \emph {et~al.}(2018)\citenamefont {Domino},
  \citenamefont {Glos}, \citenamefont {Ostaszewski}, \citenamefont {Pawela},\
  and\ \citenamefont {Sadowski}}]{domino2016properties}%
  \BibitemOpen
  \bibfield  {author} {\bibinfo {author} {\bibfnamefont {K.}~\bibnamefont
  {Domino}}, \bibinfo {author} {\bibfnamefont {A.}~\bibnamefont {Glos}},
  \bibinfo {author} {\bibfnamefont {M.}~\bibnamefont {Ostaszewski}}, \bibinfo
  {author} {\bibfnamefont {L.}~\bibnamefont {Pawela}}, \ and\ \bibinfo {author}
  {\bibfnamefont {P.}~\bibnamefont {Sadowski}},\ }\href@noop {} {\bibfield
  {journal} {\bibinfo  {journal} {Quantum Information and Computation}\
  }\textbf {\bibinfo {volume} {18}},\ \bibinfo {pages} {0181} (\bibinfo {year}
  {2018})}\BibitemShut {NoStop}%
\bibitem [{\citenamefont {Glos}\ \emph {et~al.}(2018)\citenamefont {Glos},
  \citenamefont {Miszczak},\ and\ \citenamefont {Ostaszewski}}]{glos2017limit}%
  \BibitemOpen
  \bibfield  {author} {\bibinfo {author} {\bibfnamefont {A.}~\bibnamefont
  {Glos}}, \bibinfo {author} {\bibfnamefont {J.}~\bibnamefont {Miszczak}}, \
  and\ \bibinfo {author} {\bibfnamefont {M.}~\bibnamefont {Ostaszewski}},\
  }\href {\doibase 10.1088/1751-8121/aa9a4a} {\bibfield  {journal} {\bibinfo
  {journal} {J. Phys. A: Math. Theor.}\ }\textbf {\bibinfo {volume} {51}},\
  \bibinfo {pages} {035304} (\bibinfo {year} {2018})},\ \bibinfo {note}
  {arXiv:1703.01792}\BibitemShut {NoStop}%
\bibitem [{\citenamefont {Liu}\ and\ \citenamefont
  {Balu}(2017)}]{liu_continuous-time_2016}%
  \BibitemOpen
  \bibfield  {author} {\bibinfo {author} {\bibfnamefont {C.}~\bibnamefont
  {Liu}}\ and\ \bibinfo {author} {\bibfnamefont {R.}~\bibnamefont {Balu}},\
  }\href {\doibase 10.1007/s11128-017-1625-8} {\bibfield  {journal} {\bibinfo
  {journal} {Quantum Inf. Process.}\ }\textbf {\bibinfo {volume} {16}},\
  \bibinfo {pages} {173} (\bibinfo {year} {2017})}\BibitemShut {NoStop}%
\bibitem [{\citenamefont {Domino}\ \emph {et~al.}(2017)\citenamefont {Domino},
  \citenamefont {Glos},\ and\ \citenamefont
  {Ostaszewski}}]{domino2017superdiffusive}%
  \BibitemOpen
  \bibfield  {author} {\bibinfo {author} {\bibfnamefont {K.}~\bibnamefont
  {Domino}}, \bibinfo {author} {\bibfnamefont {A.}~\bibnamefont {Glos}}, \ and\
  \bibinfo {author} {\bibfnamefont {M.}~\bibnamefont {Ostaszewski}},\
  }\href@noop {} {\bibfield  {journal} {\bibinfo  {journal} {Quantum Inform.
  Comput}\ }\textbf {\bibinfo {volume} {17}},\ \bibinfo {pages} {973} (\bibinfo
  {year} {2017})}\BibitemShut {NoStop}%
\bibitem [{\citenamefont {Bringuier}(2017)}]{bringuier_central_2016}%
  \BibitemOpen
  \bibfield  {author} {\bibinfo {author} {\bibfnamefont {H.}~\bibnamefont
  {Bringuier}},\ }\href {\doibase 10.1007/s00023-017-0597-7} {\bibfield
  {journal} {\bibinfo  {journal} {Annales Henri Poincar{\'e}}\ }\textbf
  {\bibinfo {volume} {18}},\ \bibinfo {pages} {3167} (\bibinfo {year}
  {2017})}\BibitemShut {NoStop}%
\bibitem [{\citenamefont {S{\'a}nchez-Burillo}\ \emph
  {et~al.}(2012)\citenamefont {S{\'a}nchez-Burillo}, \citenamefont {Duch},
  \citenamefont {G{\'o}mez-Gardenes},\ and\ \citenamefont
  {Zueco}}]{sanchez-burillo_quantum_2012}%
  \BibitemOpen
  \bibfield  {author} {\bibinfo {author} {\bibfnamefont {E.}~\bibnamefont
  {S{\'a}nchez-Burillo}}, \bibinfo {author} {\bibfnamefont {J.}~\bibnamefont
  {Duch}}, \bibinfo {author} {\bibfnamefont {J.}~\bibnamefont
  {G{\'o}mez-Gardenes}}, \ and\ \bibinfo {author} {\bibfnamefont
  {D.}~\bibnamefont {Zueco}},\ }\href {\doibase 10.1038/srep00605} {\bibfield
  {journal} {\bibinfo  {journal} {Sci. Rep.}\ }\textbf {\bibinfo {volume} {2}}
  (\bibinfo {year} {2012}),\ 10.1038/srep00605}\BibitemShut {NoStop}%
\bibitem [{\citenamefont {Falloon}\ \emph {et~al.}(2017)\citenamefont
  {Falloon}, \citenamefont {Rodriguez},\ and\ \citenamefont
  {Wang}}]{falloon2017qswalk}%
  \BibitemOpen
  \bibfield  {author} {\bibinfo {author} {\bibfnamefont {P.}~\bibnamefont
  {Falloon}}, \bibinfo {author} {\bibfnamefont {J.}~\bibnamefont {Rodriguez}},
  \ and\ \bibinfo {author} {\bibfnamefont {J.}~\bibnamefont {Wang}},\ }\href
  {\doibase 10.1016/j.cpc.2017.03.014} {\bibfield  {journal} {\bibinfo
  {journal} {Comput. Phys. Commun.}\ } (\bibinfo {year} {2017}),\
  10.1016/j.cpc.2017.03.014}\BibitemShut {NoStop}%
\bibitem [{\citenamefont {Loke}\ \emph {et~al.}(2017)\citenamefont {Loke},
  \citenamefont {Tang}, \citenamefont {Rodriguez}, \citenamefont {Small},\ and\
  \citenamefont {Wang}}]{loke2017comparing}%
  \BibitemOpen
  \bibfield  {author} {\bibinfo {author} {\bibfnamefont {T.}~\bibnamefont
  {Loke}}, \bibinfo {author} {\bibfnamefont {J.}~\bibnamefont {Tang}}, \bibinfo
  {author} {\bibfnamefont {J.}~\bibnamefont {Rodriguez}}, \bibinfo {author}
  {\bibfnamefont {M.}~\bibnamefont {Small}}, \ and\ \bibinfo {author}
  {\bibfnamefont {J.}~\bibnamefont {Wang}},\ }\href {\doibase
  10.1007/s11128-016-1456-z} {\bibfield  {journal} {\bibinfo  {journal}
  {Quantum Inf. Process.}\ }\textbf {\bibinfo {volume} {16}},\ \bibinfo {pages}
  {25} (\bibinfo {year} {2017})}\BibitemShut {NoStop}%
\bibitem [{\citenamefont {Caruso}\ \emph {et~al.}(2009)\citenamefont {Caruso},
  \citenamefont {Chin}, \citenamefont {Datta}, \citenamefont {Huelga},\ and\
  \citenamefont {Plenio}}]{caruso2009highly}%
  \BibitemOpen
  \bibfield  {author} {\bibinfo {author} {\bibfnamefont {F.}~\bibnamefont
  {Caruso}}, \bibinfo {author} {\bibfnamefont {A.}~\bibnamefont {Chin}},
  \bibinfo {author} {\bibfnamefont {A.}~\bibnamefont {Datta}}, \bibinfo
  {author} {\bibfnamefont {S.}~\bibnamefont {Huelga}}, \ and\ \bibinfo {author}
  {\bibfnamefont {M.}~\bibnamefont {Plenio}},\ }\href {\doibase
  10.1063/1.3223548} {\bibfield  {journal} {\bibinfo  {journal} {J. Chem.
  Phys.}\ }\textbf {\bibinfo {volume} {131}},\ \bibinfo {pages} {09B612}
  (\bibinfo {year} {2009})}\BibitemShut {NoStop}%
\bibitem [{\citenamefont {Mohseni}\ \emph {et~al.}(2008)\citenamefont
  {Mohseni}, \citenamefont {Rebentrost}, \citenamefont {Lloyd},\ and\
  \citenamefont {Aspuru-Guzik}}]{mohseni2008environment}%
  \BibitemOpen
  \bibfield  {author} {\bibinfo {author} {\bibfnamefont {M.}~\bibnamefont
  {Mohseni}}, \bibinfo {author} {\bibfnamefont {P.}~\bibnamefont {Rebentrost}},
  \bibinfo {author} {\bibfnamefont {S.}~\bibnamefont {Lloyd}}, \ and\ \bibinfo
  {author} {\bibfnamefont {A.}~\bibnamefont {Aspuru-Guzik}},\ }\href {\doibase
  10.1063/1.3002335} {\bibfield  {journal} {\bibinfo  {journal} {J. Chem.
  Phys.}\ }\textbf {\bibinfo {volume} {129}},\ \bibinfo {pages} {11B603}
  (\bibinfo {year} {2008})}\BibitemShut {NoStop}%
\bibitem [{\citenamefont {Rebentrost}\ \emph {et~al.}(2009)\citenamefont
  {Rebentrost}, \citenamefont {Mohseni}, \citenamefont {Kassal}, \citenamefont
  {Lloyd},\ and\ \citenamefont {Aspuru-Guzik}}]{rebentrost2009environment}%
  \BibitemOpen
  \bibfield  {author} {\bibinfo {author} {\bibfnamefont {P.}~\bibnamefont
  {Rebentrost}}, \bibinfo {author} {\bibfnamefont {M.}~\bibnamefont {Mohseni}},
  \bibinfo {author} {\bibfnamefont {I.}~\bibnamefont {Kassal}}, \bibinfo
  {author} {\bibfnamefont {S.}~\bibnamefont {Lloyd}}, \ and\ \bibinfo {author}
  {\bibfnamefont {A.}~\bibnamefont {Aspuru-Guzik}},\ }\href {\doibase
  10.1088/1367-2630/11/3/033003} {\bibfield  {journal} {\bibinfo  {journal}
  {New J. Phys.}\ }\textbf {\bibinfo {volume} {11}},\ \bibinfo {pages} {033003}
  (\bibinfo {year} {2009})}\BibitemShut {NoStop}%
\bibitem [{\citenamefont {Whitfield}\ \emph {et~al.}(2010)\citenamefont
  {Whitfield}, \citenamefont {Rodr{\'\i}guez-Rosario},\ and\ \citenamefont
  {Aspuru-Guzik}}]{whitfield2010quantum}%
  \BibitemOpen
  \bibfield  {author} {\bibinfo {author} {\bibfnamefont {J.}~\bibnamefont
  {Whitfield}}, \bibinfo {author} {\bibfnamefont {C.}~\bibnamefont
  {Rodr{\'\i}guez-Rosario}}, \ and\ \bibinfo {author} {\bibfnamefont
  {A.}~\bibnamefont {Aspuru-Guzik}},\ }\href {\doibase
  10.1103/PhysRevA.81.022323} {\bibfield  {journal} {\bibinfo  {journal} {Phys.
  Rev. A}\ }\textbf {\bibinfo {volume} {81}},\ \bibinfo {pages} {022323}
  (\bibinfo {year} {2010})}\BibitemShut {NoStop}%
\bibitem [{qua(012 )}]{quantiki-list}%
  \BibitemOpen
  \href@noop {} {\enquote {\bibinfo {title} {{List of QC simulators}},}\ }
  (\bibinfo {year} {2012-}),\ \bibinfo {note}
  {\url{https://quantiki.org/wiki/list-qc-simulators} (Accessed on
  18/12/2017)}\BibitemShut {NoStop}%
\bibitem [{\citenamefont {Marquezino}\ and\ \citenamefont
  {Portugal}(2008)}]{marquezino2008qwalk}%
  \BibitemOpen
  \bibfield  {author} {\bibinfo {author} {\bibfnamefont {F.}~\bibnamefont
  {Marquezino}}\ and\ \bibinfo {author} {\bibfnamefont {R.}~\bibnamefont
  {Portugal}},\ }\href {\doibase 10.1016/j.cpc.2008.02.019} {\bibfield
  {journal} {\bibinfo  {journal} {Comput. Phys. Commun.}\ }\textbf {\bibinfo
  {volume} {179}},\ \bibinfo {pages} {359} (\bibinfo {year}
  {2008})}\BibitemShut {NoStop}%
\bibitem [{\citenamefont {Berry}\ \emph {et~al.}(2011)\citenamefont {Berry},
  \citenamefont {Bourke},\ and\ \citenamefont {Wang}}]{berry2011qwviz}%
  \BibitemOpen
  \bibfield  {author} {\bibinfo {author} {\bibfnamefont {S.}~\bibnamefont
  {Berry}}, \bibinfo {author} {\bibfnamefont {P.}~\bibnamefont {Bourke}}, \
  and\ \bibinfo {author} {\bibfnamefont {J.}~\bibnamefont {Wang}},\ }\href
  {\doibase 10.1016/j.cpc.2011.06.002} {\bibfield  {journal} {\bibinfo
  {journal} {Comput. Phys. Commun.}\ }\textbf {\bibinfo {volume} {182}},\
  \bibinfo {pages} {2295} (\bibinfo {year} {2011})}\BibitemShut {NoStop}%
\bibitem [{\citenamefont {Izaac}\ and\ \citenamefont
  {Wang}(2015)}]{izaac2015pyctqw}%
  \BibitemOpen
  \bibfield  {author} {\bibinfo {author} {\bibfnamefont {J.}~\bibnamefont
  {Izaac}}\ and\ \bibinfo {author} {\bibfnamefont {J.}~\bibnamefont {Wang}},\
  }\href {\doibase 10.1016/j.cpc.2014.09.011} {\bibfield  {journal} {\bibinfo
  {journal} {Comput. Phys. Commun.}\ }\textbf {\bibinfo {volume} {186}},\
  \bibinfo {pages} {81} (\bibinfo {year} {2015})}\BibitemShut {NoStop}%
\bibitem [{\citenamefont {Schirmer}\ and\ \citenamefont
  {Wang}(2010)}]{schirmer2010stabilizing}%
  \BibitemOpen
  \bibfield  {author} {\bibinfo {author} {\bibfnamefont {S.}~\bibnamefont
  {Schirmer}}\ and\ \bibinfo {author} {\bibfnamefont {X.}~\bibnamefont
  {Wang}},\ }\href@noop {} {\bibfield  {journal} {\bibinfo  {journal} {Physical
  Review A}\ }\textbf {\bibinfo {volume} {81}},\ \bibinfo {pages} {062306}
  (\bibinfo {year} {2010})}\BibitemShut {NoStop}%
\bibitem [{\citenamefont {Gorini}\ \emph {et~al.}(1976)\citenamefont {Gorini},
  \citenamefont {Kossakowski},\ and\ \citenamefont
  {Sudarshan}}]{gorini1976completely}%
  \BibitemOpen
  \bibfield  {author} {\bibinfo {author} {\bibfnamefont {V.}~\bibnamefont
  {Gorini}}, \bibinfo {author} {\bibfnamefont {A.}~\bibnamefont {Kossakowski}},
  \ and\ \bibinfo {author} {\bibfnamefont {E.}~\bibnamefont {Sudarshan}},\
  }\href {\doibase 10.1063/1.522979} {\bibfield  {journal} {\bibinfo  {journal}
  {J. Math. Phys.}\ }\textbf {\bibinfo {volume} {17}},\ \bibinfo {pages} {821}
  (\bibinfo {year} {1976})}\BibitemShut {NoStop}%
\bibitem [{\citenamefont {Kossakowski}(1972)}]{kossakowski1972quantum}%
  \BibitemOpen
  \bibfield  {author} {\bibinfo {author} {\bibfnamefont {A.}~\bibnamefont
  {Kossakowski}},\ }\href {\doibase 10.1016/0034-4877(72)90010-9} {\bibfield
  {journal} {\bibinfo  {journal} {Rep. Math. Phys.}\ }\textbf {\bibinfo
  {volume} {3}},\ \bibinfo {pages} {247} (\bibinfo {year} {1972})}\BibitemShut
  {NoStop}%
\bibitem [{\citenamefont {Lindblad}(1976)}]{lindblad1976generators}%
  \BibitemOpen
  \bibfield  {author} {\bibinfo {author} {\bibfnamefont {G.}~\bibnamefont
  {Lindblad}},\ }\href {\doibase 10.1007/BF01608499} {\bibfield  {journal}
  {\bibinfo  {journal} {Commun. Math. Phys.}\ }\textbf {\bibinfo {volume}
  {48}},\ \bibinfo {pages} {119} (\bibinfo {year} {1976})}\BibitemShut
  {NoStop}%
\bibitem [{\citenamefont {Miszczak}(2011)}]{miszczak2011singular}%
  \BibitemOpen
  \bibfield  {author} {\bibinfo {author} {\bibfnamefont {J.}~\bibnamefont
  {Miszczak}},\ }\href {\doibase 10.1142/S0129183111016683} {\bibfield
  {journal} {\bibinfo  {journal} {Int. J. Mod. Phys. C}\ }\textbf {\bibinfo
  {volume} {22}},\ \bibinfo {pages} {897} (\bibinfo {year} {2011})}\BibitemShut
  {NoStop}%
\bibitem [{exp(2018)}]{expokit}%
  \BibitemOpen
  \href@noop {} {\enquote {\bibinfo {title} {Julia implementations of some
  routines contained in {EXPOKIT}},}\ } (\bibinfo {year} {2018}),\ \bibinfo
  {note} {\url{https://github.com/acroy/Expokit.jl} (Accessed on
  18/12/2017)}\BibitemShut {NoStop}%
\bibitem [{\citenamefont {Sidje}(1998)}]{sidje1998expokit}%
  \BibitemOpen
  \bibfield  {author} {\bibinfo {author} {\bibfnamefont {R.}~\bibnamefont
  {Sidje}},\ }\href {\doibase 10.1145/285861.285868} {\bibfield  {journal}
  {\bibinfo  {journal} {ACM Trans. Math. Software}\ }\textbf {\bibinfo {volume}
  {24}},\ \bibinfo {pages} {130} (\bibinfo {year} {1998})}\BibitemShut
  {NoStop}%
\bibitem [{\citenamefont {Bromberger}\ \emph {et~al.}(2017)\citenamefont
  {Bromberger}, \citenamefont {Fairbanks},\ and\ \citenamefont
  {et~al.}}]{lightgraphs}%
  \BibitemOpen
  \bibfield  {author} {\bibinfo {author} {\bibfnamefont {S.}~\bibnamefont
  {Bromberger}}, \bibinfo {author} {\bibfnamefont {J.}~\bibnamefont
  {Fairbanks}}, \ and\ \bibinfo {author} {\bibnamefont {et~al.}},\ }\href
  {\doibase 10.5281/zenodo.889971} {\enquote {\bibinfo {title}
  {{JuliaGraphs/LightGraphs.jl: LightGraphs}},}\ } (\bibinfo {year}
  {2017})\BibitemShut {NoStop}%
\bibitem [{pyp(2017)}]{pyplot}%
  \BibitemOpen
  \href@noop {} {\enquote {\bibinfo {title} {{PyPlot.jl} -- plotting for
  {Julia} based on matplotlib.pyplot},}\ } (\bibinfo {year} {2017}),\ \bibinfo
  {note} {\url{https://github.com/JuliaPy/PyPlot.jl} (Accessed on
  18/12/2017)}\BibitemShut {NoStop}%
\bibitem [{\citenamefont {Albeverio}\ and\ \citenamefont
  {H{\o}egh-Krohn}(1978)}]{albeverio1978frobenius}%
  \BibitemOpen
  \bibfield  {author} {\bibinfo {author} {\bibfnamefont {S.}~\bibnamefont
  {Albeverio}}\ and\ \bibinfo {author} {\bibfnamefont {R.}~\bibnamefont
  {H{\o}egh-Krohn}},\ }\href {\doibase 10.1007/BF01940763} {\bibfield
  {journal} {\bibinfo  {journal} {Commun. Math. Phys.}\ }\textbf {\bibinfo
  {volume} {64}},\ \bibinfo {pages} {83} (\bibinfo {year} {1978})}\BibitemShut
  {NoStop}%
\end{thebibliography}%


\begin{thebibliography}{0}
		\bibitem{domino2017superdiffusive_summary}
		K.~Domino, A.~Glos, M.~Ostaszewski, Superdiffusive quantum stochastic walk
		definable of arbitrary directed graph, Quantum Inform. Comput 17~(11-12)
		(2017) 973--986.
	\end{thebibliography}
%%%%%%%%%%%%%%%%%%%%%%%%%%%%%%%%%%%%%%%%%%%%%%%%%%%%%%%%%%%%%%%%%%%%%%%%%%%%%%%%

\end{document}